\documentclass[reprint,amsmath,amssymb,pra,aps, nofootinbib,superscriptaddress]{revtex4-1}

\usepackage[T1]{fontenc}
\usepackage[utf8]{inputenc}
\usepackage{amsmath, amssymb}
\usepackage{bbold}
\usepackage{bm}
\usepackage{tipa}
\usepackage{blindtext}
\usepackage[dvipsnames]{xcolor}
\definecolor{dcyan}{RGB}{0,100,100}
\usepackage{graphicx}
\usepackage{tabularx}
\usepackage{textgreek}
\usepackage{xfrac}
\usepackage{units}
\usepackage{comment}
\usepackage{float}

\usepackage{hyperref}
\usepackage{xcolor}
\definecolor{green_cust}{RGB}{0,154,85}
\definecolor{red_cust}{RGB}{173,49,54}
\definecolor{blue_cust}{RGB}{0,103,148}
\hypersetup{colorlinks=true, linkcolor=red_cust, citecolor=green_cust, filecolor=blue_cust, urlcolor=blue_cust}

\usepackage{soul}

\newcommand{\ket}[1]{|{#1}\rangle}

\newcommand{\Figref}[1]{Fig.~\hyperref[#1]{\ref{#1}}}
\newcommand{\circlenum}[1]{\raisebox{.5pt}{\textcircled{\raisebox{-.9pt} {#1}}}}

\begin{document}
\title{Disorder-Assisted Assembly of Strongly Correlated Fluids of Light}
\author{Brendan Saxberg$^*$}
\author{Andrei Vrajitoarea$^*$}
\author{Gabrielle Roberts$^*$}
\author{Margaret G. Panetta}
\affiliation{The Department of Physics, The James Franck Institute, and The Pritzker School of Molecular Engineering, The University of Chicago, Chicago, IL}
\author{Jonathan Simon}
\affiliation{The Department of Physics, The James Franck Institute, and The Pritzker School of Molecular Engineering, The University of Chicago, Chicago, IL}
\affiliation{The Department of Physics, Stanford University, Stanford, CA}
\affiliation{The Department of Applied Physics, Stanford University, Stanford, CA}
\author{David I. Schuster}
\affiliation{The Department of Physics, The James Franck Institute, and The Pritzker School of Molecular Engineering, The University of Chicago, Chicago, IL}
\affiliation{The Department of Applied Physics, Stanford University, Stanford, CA}
\date{\today}


\begin{abstract}

Guiding many-body systems to desired states is a central challenge of modern quantum science, with applications from quantum computation~\cite{laflamme1996perfect,devitt2013quantum} to many-body physics~\cite{carusotto2020photonic} and quantum-enhanced metrology~\cite{pezze2018quantum}. Approaches to solving this problem include step-by-step assembly~\cite{grusdt2014topological,gomes2012designer,dallaire2019low}, reservoir engineering to irreversibly pump towards a target state~\cite{poyatos1996quantum,kapit2014induced,lebreuilly2017stabilizing}, and adiabatic evolution from a known initial state~\cite{albash2018adiabatic,zurek2005dynamics}.  Here we construct low-entropy quantum fluids of light in a Bose Hubbard circuit by combining particle-by-particle assembly and adiabatic preparation. We inject individual photons into a disordered lattice where the eigenstates are known \& localized, then adiabatically remove this disorder, allowing quantum fluctuations to melt the photons into a fluid. Using our platform~\cite{Ma2019AuthorPhotons}, we first benchmark this lattice melting technique by building and characterizing arbitrary single-particle-in-a-box states, then assemble multi-particle strongly correlated fluids. Inter-site entanglement measurements performed through single-site tomography indicate that the particles in the fluid delocalize, while two-body density correlation measurements demonstrate that they also avoid one another, revealing Friedel oscillations characteristic of a Tonks-Girardeau gas~\cite{tonks1936complete,girardeau1960relationship}. This work opens new possibilities for preparation of topological and otherwise exotic phases of synthetic matter~\cite{goldman2016topological,carusotto2020photonic,ozawa2019topological}.
\end{abstract}


\maketitle

\section{Introduction}
Synthetic materials, which are composed of interacting ions~\cite{blatt2012quantum}, atoms~\cite{bloch2012quantum} or photons~\cite{carusotto2020photonic,clark2020observation}, rather than interacting electrons as in solid state materials, offer a unique window into the equilibrium and dynamical properties of many-body quantum systems. Near-equilibrium, minimal realizations of superconductors~\cite{chen2005bcs}, Mott insulators~\cite{greiner2002quantum}, and topological bands~\cite{ozawa2019topological,cooper2019topological} have elucidated the essential physics of these materials. Lattice-site~\cite{bakr2009quantum} and time~\cite{trotzky2008time} resolved probes have exposed previously inaccessible quantities like entanglement~\cite{islam2015measuring,karamlou2022quantum} to direct observation.
 
Recently, synthetic matter efforts have begun to explore explicitly out-of-equilibrium phenomena including time crystallinity~\cite{zhang2017observation,choi2017observation}, many-body localization~\cite{choi2016exploring,roushan2017spectroscopic}, quantum scarring~\cite{bluvstein2021controlling}, and bad-metal transport~\cite{brown2019bad}, with prospects to explore phenomena including light-induced superconductivity~\cite{mciver2020light} and measurement-induced phase transitions~\cite{choi2020quantum}. These experiments are particularly impactful for benchmarking computational tools, as late-time dynamics of moderately sized quantum systems are already beyond the capabilities of state-of-the-art numerics~\cite{eisert2015quantum}.

Often neglected is the fact that preparing equilibrium states is itself an intrinsically non-equilibrium process, because driving a quantum many-body system to a desired target state requires \emph{dynamics}. This challenge is typically overlooked in the solid state, where thermalization of long-lived electrons with broadband thermal reservoirs allows for robust entropy removal despite fundamentally inefficient thermalization. In synthetic materials, limited particle lifetimes make it crucial to develop optimized state preparation schemes that approach the fundamental quantum speed limits.

Efficient preparation of many-body states of ultracold atoms typically involves: (i) Laser cooling~\cite{phillips1998nobel}, where scattered light removes entropy from individual atoms; (ii) Evaporative cooling, where collisions dump entropy into consequently-lost atoms, producing a Bose-Einstein condensate (BEC)~\cite{ketterle2002nobel}; and (iii) Adiabatic variation of the Hamiltonian, so that the weakly interacting (BEC) ground state of the initial Hamiltonian evolves into the strongly interacting ground state of the final Hamiltonian. This approach has been employed to produce, for example, Mott insulating~\cite{greiner2002quantum} and magnetically ordered~\cite{simon2011quantum,mazurenko2017cold} synthetic matter.

Materials composed of microwave photons~\cite{carusotto2020photonic} offer the unique possibility of efficient thermalization via coupling to arbitrarily-designed low-entropy reservoirs shaped through resonant filters~\cite{Ma2017AutonomousSolids}. This approach has been employed to stabilize Mott states of light~\cite{Ma2019AuthorPhotons}, with prospects for Devil's staircase~\cite{bak1982commensurate} and Laughlin-like~\cite{umucalilar2021autonomous} matter. Such reservoir engineering works extremely well to stabilize incompressible matter; preparation of compressible phases like superfluids~\cite{gemelke2009situ} and certain quantum spin liquids~\cite{zhou2017quantum} requires new approaches.

\begin{figure*}
	\centering
	\includegraphics[width=1\textwidth]{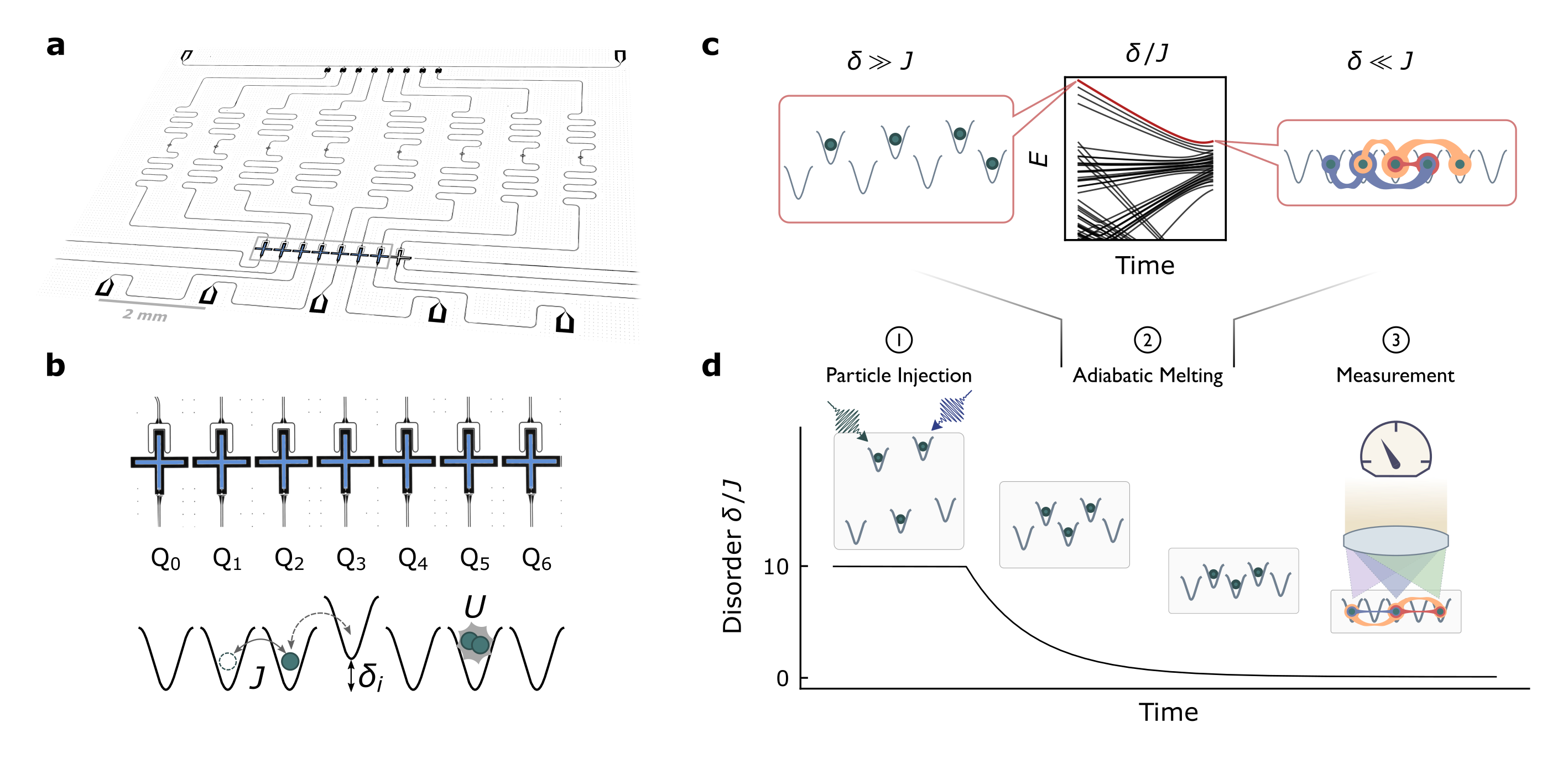}
	\caption{
		\textbf{Individually addressed many-body states in the Bose-Hubbard circuit}. The physical system, shown in \textbf{a}, consists of a one-dimensional array of capacitively coupled transmon qubits patterned on a large-area superconducting microwave circuit. This system behaves as a tight-binding lattice for photons~\cite{Ma2019AuthorPhotons} with site-resolved readout performed via microwave resonators dispersively coupled to each qubit, and real-time tuning of lattice disorder controlled by inductively coupled flux bias lines. \textbf{b.}~The physics of this 1D circuit is well characterized by the Bose-Hubbard Hamiltonian describing the dynamics of interacting particles on a lattice. The transmon qubits (highlighted in blue) realize the lattice sites in which the photonic particles reside~\cite{Koch2007}, with inter-site tunneling $J$ arising from their capacitive coupling, and the on-site interaction $U$ stemming from their anharmonicity. The flux bias lines tune the transmon energies to provide site-resolved control over lattice disorder $\delta_i$. \textbf{c,d.}~To prepare (near) arbitrary eigenstates of the disorder-free Hubbard lattice, we impose strong ($|\delta_i|\gg |J|$) controlled disorder, ensuring that all eigenstates are products of localized photons on individual sites. \circlenum{1} In this disordered configuration it is then straightforward to excite an arbitrary eigenstate by injecting photons into individual lattice sites. \circlenum{2} If the disorder is slowly removed ($|\delta_i| \rightarrow 0$), the adiabatic theorem ensures that the system always remains in the same instantaneous many-body eigenstate, resulting in a highly entangled many-body state of the disorder-free lattice. \circlenum{3} We characterize this many-body state via site-resolved occupation and correlation measurements.
	}
		\label{fig:setupfig}
\end{figure*}

In this work we harness another strength of photonic materials platforms -- particle-resolved \emph{control}, to explore a new class of state preparation schemes compatible with compressible matter. Our approach marries the addressability of particle-by-particle injection with the robustness of adiabatic evolution, enabling us to assemble arbitrary-density fluids of strongly interacting microwave photons in our 1D Bose-Hubbard circuit~\cite{Ma2019AuthorPhotons}. We first localize all eigenstates by imposing disorder much stronger than the tunneling. We then inject individual particles into localized lattice orbitals, populating a single- or many-body eigenstate of our choosing. Finally we adiabatically remove the disorder, melting this localized eigenstate into a strongly correlated fluid via tunneling-induced quantum fluctuations. To characterize the fidelity of the preparation scheme, we introduce a reversible ramp protocol that maps diabatic and decay-induced excitations of the fluid onto localized excitations in the disordered lattice. We then characterize the fluid in two ways: (i) Two-body correlation measurements, which reveal that the photons avoid one another in the fluid phase, with a universal structure characteristic of a Tonks gas; and (ii) Inter-site entanglement measurements, via the purity of single-site density matrices, which reveal that the photons delocalize during the melt and relocalize when disorder is adiabatically re-introduced.


In what follows, we first introduce the Bose-Hubbard circuit platform and its capabilities. We then describe the disorder-localized preparation scheme and test it by assembling arbitrary single-particle quasi-momentum states. We validate adiabaticity of the scheme via the reversibility of the protocol, directly measuring entropy generation in the disordered lattice. We then apply the preparation scheme to assemble few-particle states, characterizing them using new tools which reveal that the particles simultaneously delocalize and anti-bunch, hallmarks of a strongly interacting fluid of light.

\begin{figure*} 
	\centering
 	\includegraphics[width=0.9\textwidth]{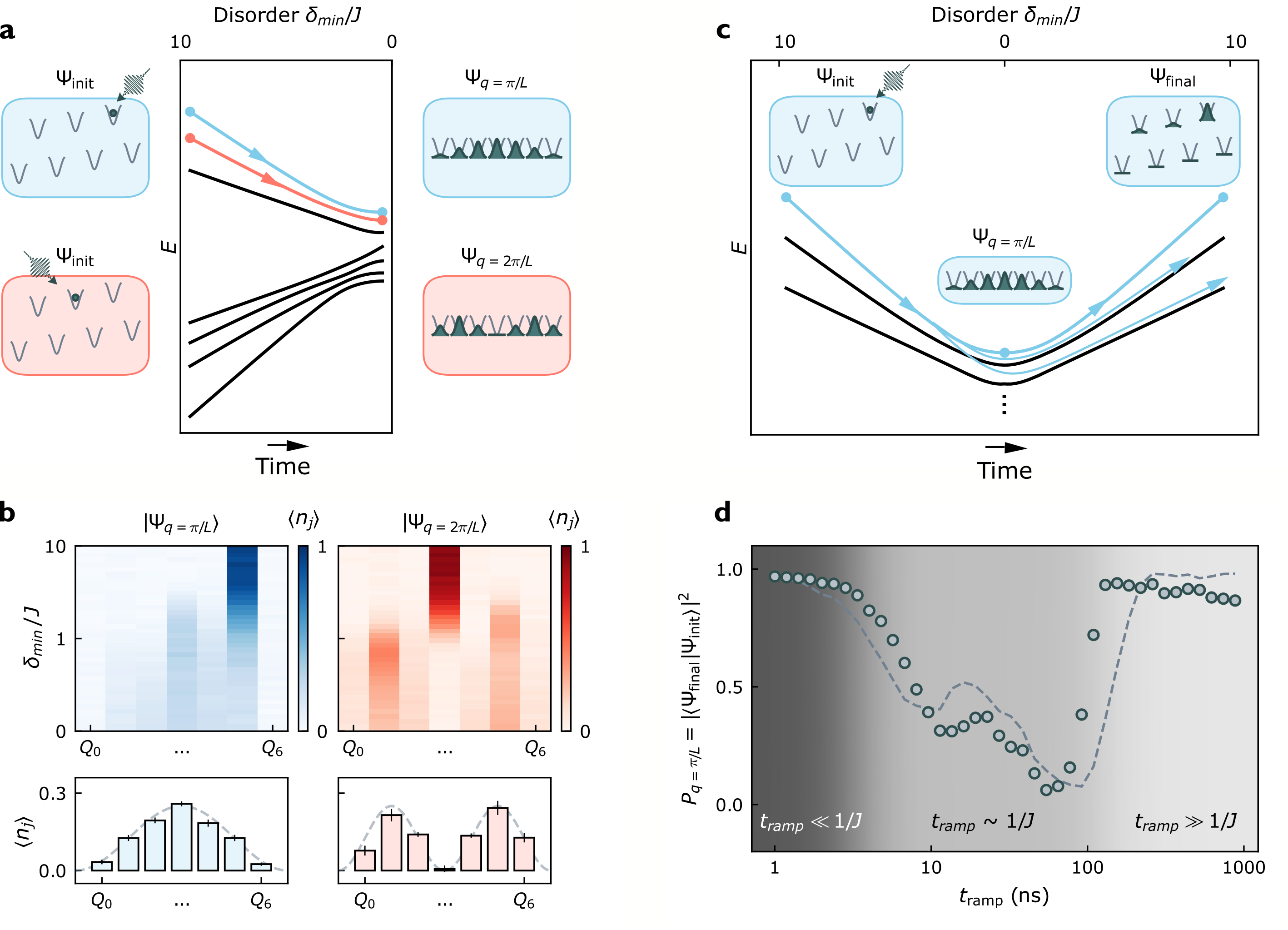}
	\caption{
		\textbf{Adiabatic assembly of single-particle eigenstates}. The simplest demonstration of the preparation protocol is the construction of single-photon particle-in-a-box states. In \textbf{a}, we plot the numerically-computed instantaneous eigenstate energies of a photon in a lattice, as disorder is reduced to zero over time. We highlight the highest (second-highest) energy eigenstates in blue (red). In the disordered lattice, the highest (second-highest) energy eigenstate is a particle localized to the single highest(second-highest) energy site, and as disorder is decreased this state is adiabatically transformed into the particle-in-a-box state with lowest (second-lowest) quasi-momentum. We demonstrate this process experimentally in \textbf{b} by assembling these lowest two quasi-momentum states. The blue (red) plots show the occupation of each of the lattice sites over time, as the disorder is adiabatically reduced to zero; here a single photon initially occupying the highest (second-highest) energy site, $Q_5$ ($Q_3$) delocalizes into the corresponding quasi-momentum state $q = \pi/L$ ($q = 2\pi/L$), shown at the final time in the bottom panels. To characterize the time required for the adiabatic sweep, we follow the protocol depicted in \textbf{c}: (i) A photon is prepared in a particular site in the presence of disorder. (ii) The disorder is ramped down and back up over a variable time $2t_{ramp}$. (iii) The final occupation of the initially prepared site $|\langle\Psi_\mathrm{final}|\Psi_\mathrm{init}\rangle|^2$ is measured. The result of this protocol for the highest energy state, shown in \textbf{d} (with dashed parameter-free theory), demonstrates that extremely fast ramps $t_\mathrm{ramp} \ll J^{-1}$ do not afford the photon sufficient time to tunnel and thus the photon remains in its initial site. At intermediate ramp speeds $t_\mathrm{ramp} \sim J^{-1}$ the photon undergoes diabatic transitions to other eigenstates and thus ends up in other lattice sites, reducing the occupation of its initial site. Only slow ramps $t_\mathrm{ramp} \gg J^{-1}$ allow the photon to adiabatically follow the initial eigenstate, delocalizing and subsequently relocalizing to its initial site. It is these slowest ramps that we employ for state preparation. Error bars reflect the S.E.M.; in \textbf{d} they are smaller than markers.
	}
	\label{fig:SingleParticleKStates}
\end{figure*} 

\section{The Bose-Hubbard circuit}

Our experiments take place in the quantum circuit shown in Fig.~\ref{fig:setupfig}a, whose physics is captured by a one-dimensional Bose-Hubbard model for photons (illustrated in Fig.~\ref{fig:setupfig}b): 
\[
\mathbf{H}_\mathrm{BH}/\hbar = -J \sum_{ \langle i,j \rangle}{a_i^\dagger a_j } + \frac{U}{2}\sum_i{n_i \left(n_i-1\right)} + \sum_i {(\omega_0+\delta_i) n_i} \label{eq:bosehubbardC}.
\]
The operator $a_i^\dagger$ ($a_i$) creates (destroys) a microwave photon on site $i$, with the number operator on site $i$ given by $n_i = a_i^\dagger a_i$. $J$ is the nearest-neighbour tunneling rate, $U$ is the on-site interaction energy, $\omega_0+\delta_i$ is the energy to create the first photon in site $i$, and $\hbar$ is the reduced Planck constant.

The lattice sites in which the photons reside are realized as transmon qubits~\cite{carusotto2020photonic}. The anharmonicity of the transmon provides the photon-photon interaction on that site. Capacitive coupling between adjacent transmons allows for nearest-neighbor tunneling, and flux loops permit qubit-by-qubit tuning of the on-site energies. We operate with $J/2\pi=9$~MHz, $U/2\pi=-230$~MHz, and site frequencies tunable in real time over $(\omega_0+\delta_i)/2\pi\in[4.1,6.1]$~GHz (see Methods). This enables introduction of disorder $\delta_i$ up to $2$~GHz, much larger than the tunneling energy. The photon lifetime in the lattice is $T_1>10\,\mu$s, so $|U|\gg J \gg 1/T_1$ providing ample time for the photons to collide, organize, and become entangled prior to decaying (see SI~\ref{SI:systemparameters}). Site-resolved microscopy is achieved by capacitive coupling of each transmon to an off-resonant coplanar waveguide resonator, enabling direct readout of each transmon's occupation number through the dispersive shift of the resonator (see SI~\ref{SI:PurcellFilter} and~\ref{SI:Readout}). Because the system operates in the hard-core limit $|U| \gg J$, the many-body states prepared are expected to be lattice-analogs of a Tonks-Girardeau gas of impenetrable bosons~\cite{Kinoshita2004ObservationGas, Carusotto2013QuantumLight}.

\section{Single-Particle Melting}

Our protocol for preparing arbitrary eigenstates of a Hubbard lattice is highlighted in Fig.~\ref{fig:setupfig}c,d. We begin by introducing strong disorder ($|\delta_i| \gg J$) in the lattice by controllably detuning the sites (see SI~\ref{SI:FluxCrosstalk}). In this disordered configuration, tunneling is suppressed and all eigenstates are products of localized photons on individual sites. Such states are easily prepared by injecting photons into individual sites with calibrated microwave \textpi-pulses. The lattice sites are then tuned into resonance by reducing the disorder $|\delta_i| \rightarrow 0$ slowly enough to maintain adiabaticity, allowing the system to remain in the same instantaneous eigenstate, which melts the particles into a correlated fluid. We then characterize the prepared states by site-resolved probes of occupation, coherence, and correlation.

We begin by applying this preparation protocol to construction of single-photon particle-in-a-box eigenstates. The dependence of these eigenstates upon disorder, and their corresponding energies, are displayed pictorially in Fig.~\ref{fig:SingleParticleKStates}a. At maximum disorder (Fig.~\ref{fig:SingleParticleKStates}a, left) each eigenstate is localized to a single site, while near zero disorder the eigenstates are delocalized particle-in-a-box states (Fig.~\ref{fig:SingleParticleKStates}a, right). Adiabatic evolution ensures that the system remains in the same instantaneous eigenstate, forming a unique connection between the lattice site the photon occupies in the disordered configuration and the state into which it delocalizes.

In Fig.~\ref{fig:SingleParticleKStates}b we prepare the highest-energy localized states and measure the evolution of their densities as we adiabatically reduce the disorder to zero. The blue (red) data show the dynamics of a single photon prepared in the highest (second-highest) energy site, $Q_5$ ($Q_3$), as it delocalizes near degeneracy. The measured density profiles of the final states are sinusoidal, with zero and one nodes, respectively, matching the probability distributions for the $q = \pi/L$ and $q = 2\pi/L$ particle-in-a-box/quasi-momentum states. Note that we quote $J>0$ in our platform, with a global minus sign for the kinetic term in the Hamiltonian. In a hard-core band with a fixed photon number, the lowest and highest-energy eigenstates are connected through a gauge transformation (by changing the sign of $J$). Without impacting the physics, we focus our efforts in preparing the higher-energy states in each particle sector.

Because photon loss renders our quantum system inherently open, we must balance the need for slow evolution (set by the energy gaps) to satisfy adiabaticity with the need to evolve faster than photon loss. We have developed a protocol to experimentally probe the adiabaticity of the process and extract the optimum ramp rate, without needing to perform tomography on a highly entangled state: We ramp the disorder down to zero and then back up, exactly reversing the downward ramp. We then measure the fraction of the time that the photon returns to its initial lattice site. Because all eigenstates are localized in the final, disordered lattice, this is a direct measure of the overlap of the final state with the initial state $|\langle \Psi_\mathrm{final}|\Psi_\mathrm{init}\rangle|^2$. Performing this experiment versus total evolution time $2 t_\mathrm{ramp}$, provides a measure of adiabaticity.

This process is depicted in Fig.~\ref{fig:SingleParticleKStates}c for a single particle, where it is applied to the preparation of the highest-energy quasi-momentum state as shown in Fig.~\ref{fig:SingleParticleKStates}d. For very fast ramps $t_\mathrm{ramp} \ll J^{-1}$, the photon remains in its initial site because it lacks the time to tunnel even to its nearest neighbor. For intermediate ramp speeds $t_\mathrm{ramp} \sim J^{-1}$, the photon has time to delocalize but not to adiabatically follow, and thus undergoes diabatic transitions to other quasi-momentum states as the disorder is reduced, leading to a decreased population of the initial site. When the ramp is sufficiently slow $t_\mathrm{ramp} \gg J^{-1}$ the photon is able to delocalize, adiabatically follow the same eigenstate, and relocalize, resulting in near-unity occupation in the initial site. We employ extensions of this technique to multiple particles (see SI~\ref{SI:AdditionalData}) behind the scenes in order to optimize performance throughout the remainder of this work. This approach is particularly powerful because it is agnostic to the details of the physical platform and target state.

\begin{figure*} 
	\centering
	\includegraphics[width=1\textwidth]{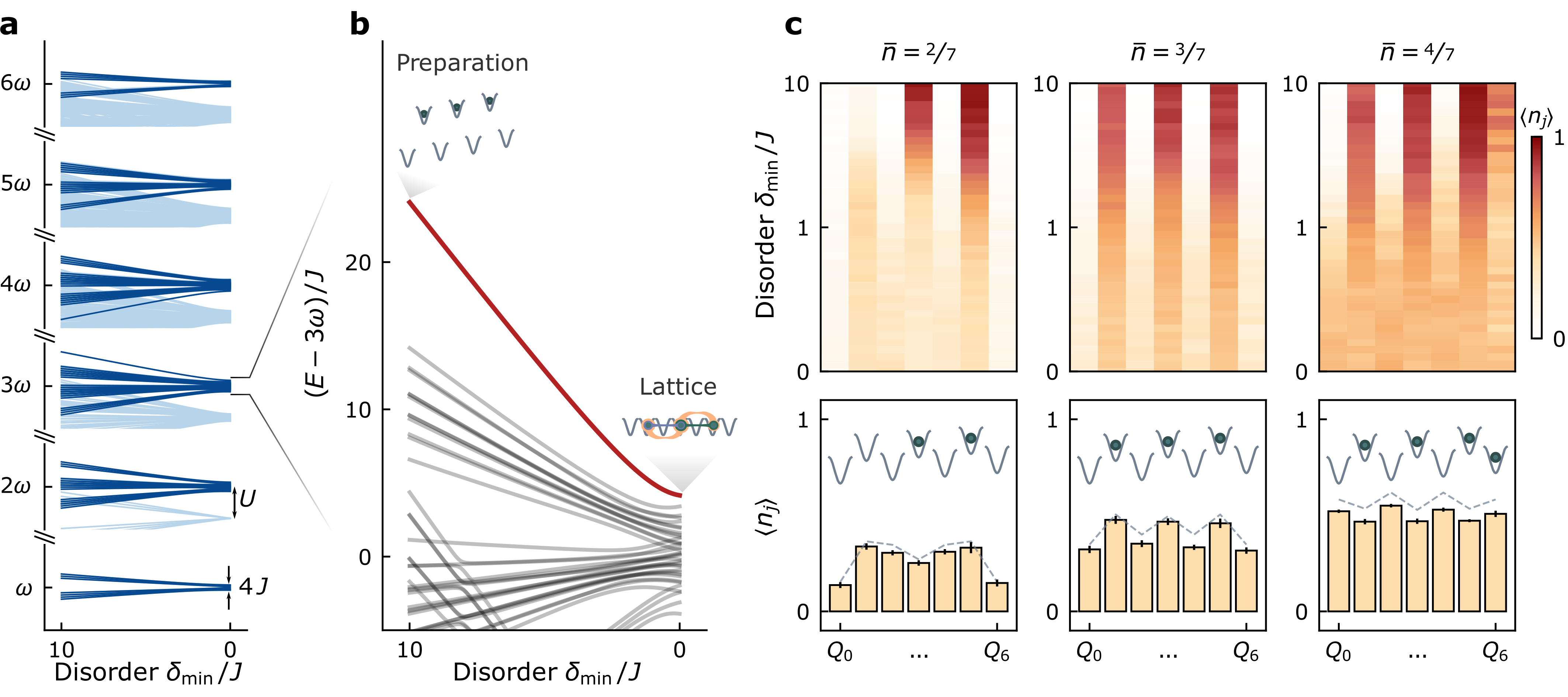}
	\caption{
		\textbf{Adiabatic preparation of strongly correlated fluids of light}. The ground states of the Hubbard lattice below unit filling are compressible fluids because motion of the particles is not fully blocked by collisions as it would be in the unit-filled Mott state. This is reflected in the many-body spectrum in \textbf{a}, where in the absence of disorder, there are bands of states (dark blue) of width $\sim J$ in which the photons do not overlap with one another, rather than a single state, as would be the case for an incompressible system. These states are spectroscopically isolated by the onsite interaction $U$ from all other states (light blue). In the incompressible Mott state there is a single gapped (by $U$) ground state. In the particular case of three particles in the lattice shown in \textbf{b}, the highest energy state, which is the fluid `ground' state (because $U,J<0$), exhibits the largest energy gap to all other states, and can thus be prepared most quickly. In \textbf{c} we adiabatically prepare the fluid ground states for two, three, and four particles in the seven site lattice. The upper panels display the density profiles for these states as the system is tuned from disordered to ordered configurations and the particles delocalize and become entangled. We highlight the measured fluid densities at degeneracy (bars in lower panels) compared to theory (dashed curve). The insets depict particle placements in the disordered configuration, which adiabatically connect to the fluids.
	}
	\label{fig:multiparticlefluid}
\end{figure*}

\section{Correlated Fluid Melting}

When multiple photons simultaneously reside in our Hubbard circuit, interactions strongly modify the behaviour of the system. At unit average occupancy ($\bar{n} \equiv N/L=1$, where $N$ is the number of photons and $L$ is the number of sites), the ground state is a Mott insulator~\cite{Ma2019AuthorPhotons} because the photons cannot move without immediately hitting a neighbor, so transport is fully impeded. Away from unit occupancy the ground state is a fluid: even if $U$ fully blocks photons passing through one another, they can still delocalize, move, and exchange momentum with their neighbors.

Strongly interacting fluids are challenging to prepare by reservoir engineering techniques~\cite{Ma2017AutonomousSolids}, which rely on irreversible photon injection that halts when adding the next photon costs substantially more energy than prior photons. The delocalization of the photons in the fluid makes it compressible: the energy required to inject additional photons changes smoothly with density, changing abruptly only at the unit-filled Mott state.

Here we prepare a compressible, strongly correlated fluid of light away from unit filling via a multi-particle variant of our disorder-assisted preparation scheme: we determine the filling by the number of photons that we coherently inject, and eigenstate by the sites into which we inject photons. In our $L=7$ site lattice we inject up to $N = 6$ photons in order to remain below unit filling. Since our interactions are attractive, $U < 0$, the `ground' state is actually the highest energy state for a given photon number; beyond this detail, the sign of $U$ does not impact the physics.

Fig.~\ref{fig:multiparticlefluid}a depicts the many-body spectrum for various fluid photon numbers versus disorder. The spectrum splits first into bands of fixed photon number separated by $\omega_0$, and then into bands of fixed numbers of overlapping particles separated by the interaction energy $U$. Within each of these bands, the states are split by fractions of the tunneling energy $J$, reflecting phonon excitations of the fluid. Fig.~\ref{fig:multiparticlefluid}b provides a detailed view of the spectrum relevant for preparing a three-photon ground state. The preparation trajectory is highlighted, beginning in the disordered lattice with photons occupying the three highest-energy sites, and ending in the ordered lattice with the photons delocalized and entangled.

We perform the disorder-assisted preparation for up to four photons, measuring, in Fig.~\ref{fig:multiparticlefluid}c, average density profiles as the lattice is tuned from disordered to disorder-free configurations (see SI~\ref{SI:AdditionalData} for $5\,\& \,6$ photons). These data demonstrate that during the melt the photons delocalize from their initial sites into all lattice sites, with the melted density profiles in good agreement with the disorder-free numerics from exact diagonalization.

Because we operate at large $U/J$, the physics is well-captured by the Tonks-Girardeau model (see SI~\ref{SI:TGgas}), whose ground state is of the Bijl-Jastrow~\cite{bijl1940lowest} form. This wavefunction is written as the product of single- and two-particle components $\Psi_B(\mathbf{x}) = \phi(\mathbf{x}) \varphi(\mathbf{x})$, for $\mathbf{x} = (x_0, x_1,\ldots, x_6)$~\cite{Rigol2011Bose1DRev}. The single-particle component $\phi(\mathbf{x}) = \prod_{i=0}^{6} \cos(\pi x_i / L)$ places each photon in the lowest-energy particle-in-a-box state of the lattice, while the two-particle component $\varphi(\mathbf{x}) = \prod_{i<j} |x_i - x_j|$ keeps the photons apart (whilst minimizing their kinetic energy) by ramping the wavefunction to zero whenever they overlap.

In the absence of interactions, the ground state density would be independent of particle number up to an overall scale, reflecting the single-particle eigenstate $q = \pi/L$ of Fig.~\ref{fig:SingleParticleKStates}b. The increasingly large deviations from such a sinusoidal form as density increases indicate that photon-photon collisions are shaping the fluid density profile. A deeper understanding of the fluid's structure requires exploring correlations and entanglement, which is the subject of the remainder of this investigation.

\begin{figure*} 
	\centering
	\includegraphics[width=0.95\textwidth]{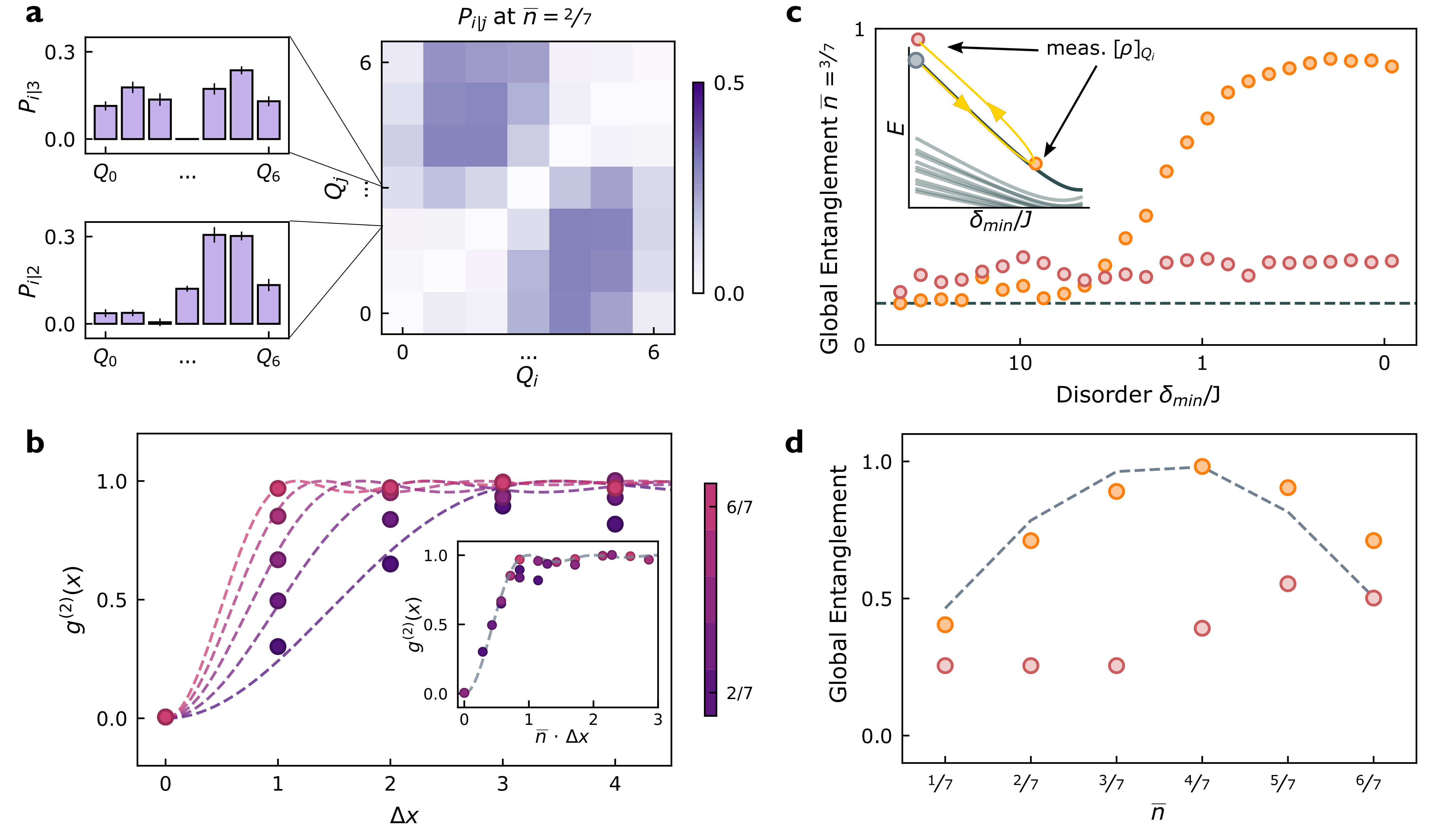}
	\caption{
		\textbf{Microscopy of the strongly correlated fluid: anti-bunching and delocalization}. The correlated photon liquid is characterized simultaneously by photons (i) avoiding one another while (ii) managing to delocalize. To show that the photons avoid one another in the fluid, in \textbf{a} we measure the two-body correlator $P_{i|j}$ at $\bar{n}=\sfrac{2}{7}$. This correlator quantifies the probability of detecting a photon in site $i$ given that one was detected in site $j$. Its strong suppression near $i=j$ reflects not only the hard-core constraint, but the additional preference of the photons to minimize their kinetic energy by reducing the curvature of their wavefunction and thus avoid nearby sites. The upper (lower) inset showing $P_{i|3}$ ($P_{i|2}$) demonstrates that when the first photon is detected in the middle site of the lattice (one site left of the middle), the second photon minimizes its energy by occupying particle-in-a-box states in two boxes on either side of the first photon created by the hardcore constraint. \textbf{b} shows the normalized two-body correlator measured for various particle numbers (densities), demonstrating that the anti-correlation length decreases as the density increases, in good agreement with a parameter-free Tonks-Girardeau theory (see SI~\ref{SI:TGgas}) and the intuition that each photon occupies less space at higher densities. Also apparent are Friedel oscillations at wavevector $k_F=\pi\bar{n}$, a signature of the fermionization of the photons. We omit large-separation correlation data, which is polluted by edge effects (see SI~\ref{SI:AdditionalData}). To probe the delocalization of the photons during the three-particle fluid preparation process, in \textbf{c} we measure the average entanglement between individual lattice sites with the rest of the system, $E_{gl}\equiv 2(1-\langle p_i\rangle)$ (orange) measured via the purity of the corresponding single-site reduced density operator $p_i\equiv Tr(\rho_i^2)$. In the disordered lattice the many-body state is a product of localized photons so the entanglement is small. In the disorder-free lattice the photons delocalize, the single-site purity drops and the entanglement increases. We rule out the possibility that this is entanglement with the \emph{environment} (decoherence) rather than with other lattice sites by ramping back to the initial configuration prior to measuring the entanglement (red). That this quantity remains small proves that the system remains un-entangled with the environment. The inset depicts the measurement time within the ramp for each curve. As shown in \textbf{d}, the entanglement peaks at half-filling because that is where knowledge of a given site's occupancy provides the most information about the rest of the system's wavefunction. Particle-hole symmetry about half-filling ($\bar{n}=\sfrac{1}{2}$) in the theory (dashed) is not reflected in the data due to the increased decay of fluids with more particles (apparent from the increased entanglement after the reverse ramp at high densities, red). Error bars represent the S.E.M. (see SI~\ref{SI:errorbars}); where absent, they are smaller than the data points.}
	\label{fig:correlations}
\end{figure*}

\section{Fluid Correlations}
\label{sec:correlations}

In the Tonks regime the photon fluid should exhibit short-range repulsion between the particles arising from the Hubbard $|U|\gg J$. We probe this physics directly through the two-body correlator $P_{i|j}$ which quantifies the probability of detecting a photon at site $i$ conditioned on one being detected at site $j$. If we consider a two-particle state with a wavefunction $\Psi(x_1, x_2)$, the detection of a particle at lattice site $x_1 = x_j$ collapses the wavefunction to a product state  $\Psi(x_1, x_2) = \delta(x_1-x_j)\Psi^\prime(x_2)$, with the conditional probability of the second photon given by $P_{i|j} = |\Psi^\prime(x_i)|^2$. This measurement is performed on a minimal two-particle fluid in Fig.~\ref{fig:correlations}a, where the insets show $P_{i|2}$ and $P_{i|3}$. The strong occupation suppression at $i=j$ reflects the hard-core constraint that forces the system wavefunction to vanish when two particles are on top of each other. Furthermore, the suppression \emph{near} $i=j$ reflects the photons' preference to minimize their wavefunction curvature and thus kinetic energy. The projective measurement of a photon in the lattice has effectively reshaped the two-particle fluid into a single-particle fluid confined in \emph{two} boxes, where the pinned (detected) photon acts as a potential barrier for the second photon.

To probe two-body correlations in fluids of more than two photons, we measure the system-averaged, normalized two-body correlator given by:
\[
g^{(2)}(x) = \frac{1}{\bar{n}^2} \sum_i{\langle n_{i}\,n_{i+x} \rangle},
\]
which quantifies the probability of simultaneously detecting two particles separated by x (in lattice sites), normalized to the average density $\bar{n}$. The central intuition, that each photon has less `space' at higher densities, is captured in Fig.~\ref{fig:correlations}b, where the antibunched region ($g^{(2)}(x)<1$) gets narrower as density increases. Indeed, when the separation is rescaled by the density, the correlator collapses onto a universal parameter-free Tonks-Girardeau theory, with each particle occupying a volume $1/\bar{n}$, and characteristic Friedel oscillations. These correlations, oscillating at the Fermi momentum, $k_F = \pi \bar{n}$, are a direct signature of the `fermionization' of the photons~\cite{girardeau1960relationship}.

\section{Fluid Delocalization \& Entanglement}
\label{sec:delocalization}
As the photons melt into a fluid they optimize their energy by delocalizing as much as possible whilst avoiding one another. The two-body correlator explored in Sec.~\ref{sec:correlations} quantifies this avoidance, and in this section we explore their delocalization. To achieve this we probe the entanglement of a single site with the remainder of the system by employing a metric developed for interacting spins~\cite{Meyer2002, Amico2008EntanglementSystems}: we measure the reduced density matrix of each individual lattice site $\rho_i$, quantifying how strongly it is entangled with rest of the lattice from its impurity $1 - \mathrm{Tr}(\rho_i^2)$. Our global measure of multipartite entanglement/delocalization is this impurity averaged over all sites:

\[
E_\text{gl} = 2-\frac{2}{N}\sum_{i=1}^N{\mathrm{Tr}(\rho_i^2)}.
\]

We wish to understand how the global entanglement scales with the number of particles and lattice disorder, which dictates the degree of delocalization. In Fig.~\ref{fig:correlations}c the entanglement is measured for a three-particle fluid as we vary the disorder along the adiabatic preparation trajectory. In the limit of strong lattice disorder the entanglement between sites is very small since the three-particle state is a product of localized photons. As we reduce the disorder, the entanglement grows, saturating to a maximum value when the sites are degenerate and the photons become fully delocalized. Residual entanglement in the disordered lattice arises from dissipative coupling to the environment -- decoherence. We exclude the possibility that decoherence is the source of entanglement in the fluid phase by ramping the lattice back to its initial disordered configuration. The fact that the measured entanglement drops definitively proves that the entanglement observed at degeneracy comes from delocalization, and not dissipation. Similar measurements and conclusions are extracted for all the other particle sectors (SI~\ref{SI:AdditionalData}). 

The dependence of the entanglement at degeneracy on the average density is displayed together with the expected theoretical calculation in Fig.~\ref{fig:correlations}d. The discrepancy at larger filling fractions is due to increased particle loss, as anticipated from the increased entanglement after the time-reversed ramp. As highlighted in theory, there is a particle-hole symmetry in the extracted entanglement measure. This is expected because in the hardcore limit, particles at filling $\bar{n}$ tunnel and avoid one another analogously to holes at filling $1-\bar{n}$. Entanglement is maximized at half-filling because that is the situation in which knowledge of the occupancy of any given site provides the \emph{most} information about the occupancy of adjacent sites: at lower (higher) fillings, most sites are empty (occupied), so knowledge of any given site's occupancy provides less information.

\section{Outlook}

In this work we have demonstrated a way to harness controlled disorder to individually index and prepare the eigenstates of a strongly interacting many-body system. In particular, we have assembled quantum fluids of light in a 1D Bose-Hubbard circuit composed of capacitively coupled transmon qubits. Leveraging our site-resolved tuning capabilities, we tune the qubits out of resonance with one another, individually excite only particular qubits, and then melt these excitations into a fluid. Site-resolved probes of correlation, entanglement, and reversibility reveal that this system realizes a Tonks-Girardeau gas~\cite{tonks1936complete,girardeau1960relationship}. It remains to be seen how this adiabatic preparation approach scales with system size; it seems certain that it will be controlled by the Kibble-Zurek mechanism~\cite{chandran2012kibble} in the thermodynamic limit, though the precise disorder employed will likely impact the structure of the excitations generated at the critical point.

By combining the techniques developed in this work with topological~\cite{roushan2017spectroscopic,owens2021chiral} circuit lattices, it should be possible to prepare topological fluids of light~\cite{grusdt2014topological}. In conjunction with auxiliary qubits, these adiabatic preparation techniques will enable direct measurements of out-of-time-order correlators \& information scrambling~\cite{swingle2016measuring}, as well as anyon statistics~\cite{grusdt2016interferometric}. 


\section{Acknowledgments}

This work was supported by ARO MURI Grant W911NF-15-1-0397, AFOSR MURI Grant FA9550-19-1-0399, and by NSF Eager Grant 1926604. Support was also provided by the Chicago MRSEC, which is funded by NSF through Grant DMR-1420709. G.R. and M.G.P acknowledge support from the NSF GRFP. A.V. acknowledges support from the MRSEC-funded Kadanoff-Rice Postdoctoral Research Fellowship. Devices were fabricated in the Pritzker Nanofabrication Facility at the University of Chicago, which receives support from Soft and Hybrid Nanotechnology Experimental (SHyNE) Resource (NSF ECCS-1542205), a node of the National Science Foundation’s National Nanotechnology Coordinated Infrastructure.

\section{Author Contributions}

The experiments were designed by B.S., A.V., G.R., J.S., and D.S. The apparatus was built by B.S., A.V., and G.R. The collection of data was handled by B.S., A.V., and G.R. All authors analyzed the data and contributed to the manuscript.

\section{Competing Interests}

The authors declare no competing financial or non-financial interests.



\clearpage
\newpage
\bibliographystyle{naturemag}
\bibliography{references}

\clearpage
\section{Methods}

To generate the Bose-Hubbard Hamiltonian for microwave photons we fabricate a one-dimensional chain of capacitively coupled transmon qubits, generating a tunneling energy $J \sim 2\pi\times 9$ MHz between the qubit lattice sites and make use of the nonlinearity of the qubits to generate interactions between photons $U \sim 2\pi\times 250$ MHz. Each transmon lattice site is capacitively coupled to a linear resonator and filter (SI~\ref{SI:PurcellFilter}), which we probe to determine the state of the connected qubit. The device is fabricated using a Ta (200~nm) base layer and Al Josephson junctions on a sapphire substrate. Detailed information on the fabrication process and system parameters can be found in SI~\ref{SI:DeviceFabandParams}. To converge to these parameters we use finite element simulation techniques, mainly HFSS, in the design phase and iterate changes based on experimental measurements.

The sample is mounted and wire-bonded to a copper PCB with a global solenoid attached, which in turn, is encased in shielding consisting of copper, MuMetal and lead, to prevent radiation interacting with the sample. The shielded sample is mounted to the mixing chamber plate of a dilution refrigerator cooling the sample below $T_{c}$ to 9~mK. Microwave coaxial cables and DC twisted pair wires feed signals from a room temperature homodyne measurement setup (Fig.~\ref{fig:SI wiring}) into the shielded sample. We correct distortions of signals sent to the qubit from the wiring and filters by inverting the transfer function and pre-applying a kernel to the signal (SI~\ref{SI:Kernel}).

By sending DC currents near the SQUID loops of the transmon qubits we bias the local magnetic field and allow local tuning frequencies $4\sim 6$~GHz, allowing us to introduce disorder $\delta_i \gg J$ to the array of transmon lattice sites, localizing the eigenstates of the system to excitations on single sites, which we selectively populate by driving power at that frequency using the common feedline. The mutual inductance between qubit SQUID loops creates cross-talk when applying current, which we address by inverting a measured crosstalk matrix and applying linear corrections to residual error (SI~\ref{SI:FluxCrosstalk}). The global solenoid field at the sample is used as an additional knob to bias the qubits initially and minimize the thermal load at the fridge needed to run experiments.

Initially our lattice starts in a configuration $|\delta_{i}| > U$, then we jump to a configuration $|\delta_i| < U$ and apply excitations to transmon lattice sites to avoid further Landau-Zener processes during state preparation (SI~\ref{SI:PulseSequencesandOperatingPoints}). We then adiabatically remove the remaining disorder to selectively generate desired compressible fluid states on the disorder-less lattice. To optimize performance of the adiabatic trajectory in state preparation we optimize reversability of the trajectory: $|\langle \psi_{i} | \psi_{f} \rangle |^2$ (Fig.~\ref{fig:SingleParticleKStates}d). 

To read out the state of the lattice we probe the response of RF signals sent down into the common feedline at the frequency of the dispersively coupled readout resonators. To extract qubit populations in the $|0\rangle$ and $|1\rangle$ states in this regime we probe at two separate frequencies that maximally distinguish the qubit states $|0\rangle$ from $|1\rangle,|2\rangle$ and $|1\rangle$ from $|0\rangle,|2\rangle$ and assume population lies within only these three states. We use a confusion matrix to correct for errors in binning, with additional errors stemming from readout crosstalk and Landau-Zener processes (SI~\ref{SI:Readout}).

We measure the assembly of single and multi-particle fluid eigenstates by measuring the population across all lattice sites, sampled across the parameterized removal of disorder $\delta_i$ (Fig.~\ref{fig:SingleParticleKStates}b,\ref{fig:multiparticlefluid}c). To characterize the physics of particle interaction in states with $n \geq 2$ photons, we measure the two-body correlator given by conditional measurement $P_{i|j}$ (Fig.~\ref{fig:correlations}c) and function $g^2(x)$ (Fig.~\ref{fig:correlations}d), where interactions influence the shape of the remaining one-particle wavefunction after measurement and the latter reveals 
variation in anti-bunching across densities. These results coincide with a parameter-free TG model (SI~\ref{SI:TGgas}). We measure delocalization of the compressible fluid states ($\bar{n} = \sfrac{1}{7} - \sfrac{6}{7}$) as disorder is removed via the global entanglement entropy. Fig.~\ref{fig:correlations}d shows our results for $\bar{n} = \sfrac{3}{7}$ while measurements and theory results for global entanglement across the adiabatic disorder sweep for all densities is displayed in Fig.~\ref{fig:SI Egl all}. 

\subsection{Data Availability}

The experimental data presented in this manuscript are available from the corresponding author upon request, due to the proprietary file formats employed in the data collection process.

\subsection{Code Availability}
The source code for simulations throughout are available from the corresponding author upon request. 

\subsection{Additional Information}
Correspondence and requests for materials should be addressed to D.S. (dis@uchicago.edu). Supplementary information is available for this paper.

\onecolumngrid

\pagebreak


\section*{Supplementary Information}
\appendix
\renewcommand{\appendixname}{Supplement}
\renewcommand{\theequation}{S\arabic{equation}}
\renewcommand{\thefigure}{S\arabic{figure}}

\setcounter{figure}{0}

\setcounter{table}{0}

\section{Device Fabrication and Parameters}
\label{SI:DeviceFabandParams}
The device is a one-dimensional chain of capacitively coupled transmons, each with $E_c/2\pi \approx 250$~MHz and flux tunable frequency of $\omega_{01}/2\pi \approx (4.0-6.0)$~GHz. The nearest-neighbor capacitive coupling is $J/2\pi\approx 9$~MHz. Next-to-nearest-neighbor tunneling is suppressed, expected to be of magnitude $\approx 0.06\times J$ from microwave finite-element simulations.

Each qubit is capacitively coupled to an individual $\lambda/2$ co-planar waveguide (CPW) readout resonator. The readout resonators are evenly staggered between $6$~GHz and $7$~GHz, with linewidths $\kappa/2\pi \approx 100$~kHz. Qubit-resonator dispersive shifts are $\chi/2\pi\approx$ 0.5-1.7 MHz, enabling high-fidelity ($>85\%$) single-shot readout. Each readout resonator is in turn capacitively coupled to a $\lambda/2$ CPW Purcell resonator, all of which are coupled to a common transmission line bus. The large-bandwidth Purcell filters allow us to drive the qubits through the common transmission line while preserving qubit lifetime. See SI~\ref{SI:PurcellFilter} for more details on the Purcell filter setup. Each qubit consists of a superconducting quantum interference device (SQUID) loop inductively coupled to a flux bias line with a mutual inductance of about 0.15 pH. The flux bias lines are used for both DC and RF tuning. Characterization of the crosstalk between flux lines is detailed below in SI~\ref{SI:FluxCrosstalk}.

The device also includes a lossy $\lambda/2$ CPW resonator and drive line coupled capacitively to the edge of the qubit chain. These circuit components are not used in this experiment; they can be used for stabilization in future experiments. Because the qubit on the edge of the chain is directly coupled to the lossy resonator, it has a lower lifetime than the other qubits. Although there are eight qubits in the chain, only seven are used, with the eighth qubit detuned to $ \omega_{01}/2\pi = 5.75$ GHz, $750$ MHz above any frequency value of any other qubits used during the experimental sequence.

See Table~\ref{SI:systemparameters} for detailed system parameters.

The sample is a $10\times 20$~mm sapphire chip with a tantalum base layer and aluminum Josephson junctions and SQUID loops. Our substrate is a 450 $\mu$m thick C-plane sapphire wafer that has been annealed at 1500\textdegree~C for 2 hours, solvent cleaned, etched in 80\textdegree C Nano-Strip\textsuperscript{\textregistered} for 10 min, and then etched again in 140\textdegree~C sulfuric acid to fully remove all contaminants~\cite{HouckTA}. The large scale features of the device are defined using optical lithograpy. The base layer is 200 nm of tantalum deposited at 800\textdegree~C, then patterned with a direct pattern writer (Heidelberg MLA 150) and wet etched in HF. Next, the junctions and SQUID loops are defined with electron beam lithography, using an MMA-PMMA bilayer resist, written on a Raith EBPG5000 Plus E-Beam Writer. The Al/AlOx/Al junctions are e-beam evaporated in an angled evaporator (Plassys MEB550). Before Al deposition, Ar ion milling is used on the exposed Ta to etch away the Ta oxide layer in order to ensure electrical contact between the Ta and Al layers. The first layer of Al (60 nm, deposited at 0.1~nm/s) is evaporated at an angle of 30\textdegree~to normal, followed by static oxidation in O\textsubscript{2} for 24 minutes at 50~mBar. The second layer of Al ($150$~nm, $0.1$~nm/s) is then evaporated at 30\textdegree~to normal but orthogonal to the first layer in the substrate plane to form the Manhattan-style junctions. The tunable transmon SQUID consists of two asymmetric square-shaped junctions with sizes of 130 nm and 230 nm, embedded in a loop of dimension 20~$\mu$m by 8~$\mu$m.

The device is then mounted and wirebonded to a multilayer copper PCB. The ground plane around the device features is also heavily wirebonded to avoid slotline modes and to fully connect the ground plane across the chip. The device is enclosed in an OFHC copper mount designed to eliminate spurious microwave modes near our operating frequencies.

\begin{center}
{\renewcommand{\arraystretch}{1.2}
\begin{table}
\begin{tabular}{|c|c|c|c|c|c|c|c|}

\hline
 \textbf{Qubit} & \textbf{1} & \textbf{2} & \textbf{3} & \textbf{4} & \textbf{5} & \textbf{6} & \textbf{7} \\
\hline
 $U_{\textrm{lattice}}/2\pi$ (MHz) & -241 & -240 & -240 & -239 & -239 & -239 & -240  \\ 
 \hline
 $J_{i, i+1}/2\pi$ (MHz) & 9.0625 & 9.032 & 8.842 & 8.936 & 9.023 & 9.040 & --  \\ 
 \hline
 $\omega_{\textrm{disorder-large}}/2\pi-4000 $ (MHz) & 410 & 992 & 466 & 956 & 513 & 1020 & 489 \\ 
 \hline
 $\omega_{\textrm{disorder-small}}/2\pi-4000 $ (MHz) & 742 & 860 & 767 & 875 & 782 & 890 & 797  \\ 
 \hline
 $T1 (\mu s)$ & 14.6 & 35.5 & 57.7 & 28.4 & 60.3 & 54.7 & 40.0  \\
 \hline
 $T2* (\mu s)$ & 0.85 & 0.64 & 1.31 & 0.77 & 3.57 & 0.84 & 1.4  \\
 \hline
 fidelity$_{\textrm{ge}}$  &0.91  &0.92  &0.93 &0.95  &0.87  &0.92  &0.83  \\
 \hline
 $\omega_{\textrm{read}}/2\pi$ (GHz) &6.197& 6.323& 6.427& 6.556& 6.655& 6.78 & 6.871 \\
 \hline
 $\kappa_{\textrm{read}}/2\pi$ (KHz) & 359 & 553 & 203 & 235 & 292 & 220&
       894 \\
 \hline
 $g_{\textrm{rd-qb}}/2\pi$ (MHz) & 60 & 63 & 72 & 64 & 78 & 70 & 70   \\
 \hline
 $\chi_{\textrm{rd-qb}}/2\pi$ (MHz) & 0.48 & 1.23 & 0.78 & 1.24 & 0.90 & 1.71 & 0.73   \\ 
 \hline
 $\omega_{\textrm{purcell}}/2\pi$ (GHz) & 6.256& 6.486& 6.706& 6.936& 7.055& 6.843& 6.604 \\
 \hline
 $\kappa_{\textrm{purcell}}/2\pi$ (MHz) & 77.5 &  52.7 &  92.5 &  72.4 & 103.1 & 56.9 & 60.8 \\
 \hline
 $g_{\textrm{rd-pf}}/2\pi$ (MHz) & 3 & 3.5 & 6 & 6.5 & 5 & 6 & 4.5 \\

\hline
\end{tabular}
\caption{
	\textbf{System Parameters} 
}
\label{SI:systemparameters}
\end{table}
}
\end{center}

\section{Microwave Measurement Setup: Cryogenic and Room Temperature Wiring}

\begin{figure*} 
	\centering
	\includegraphics[width=0.95\textwidth]{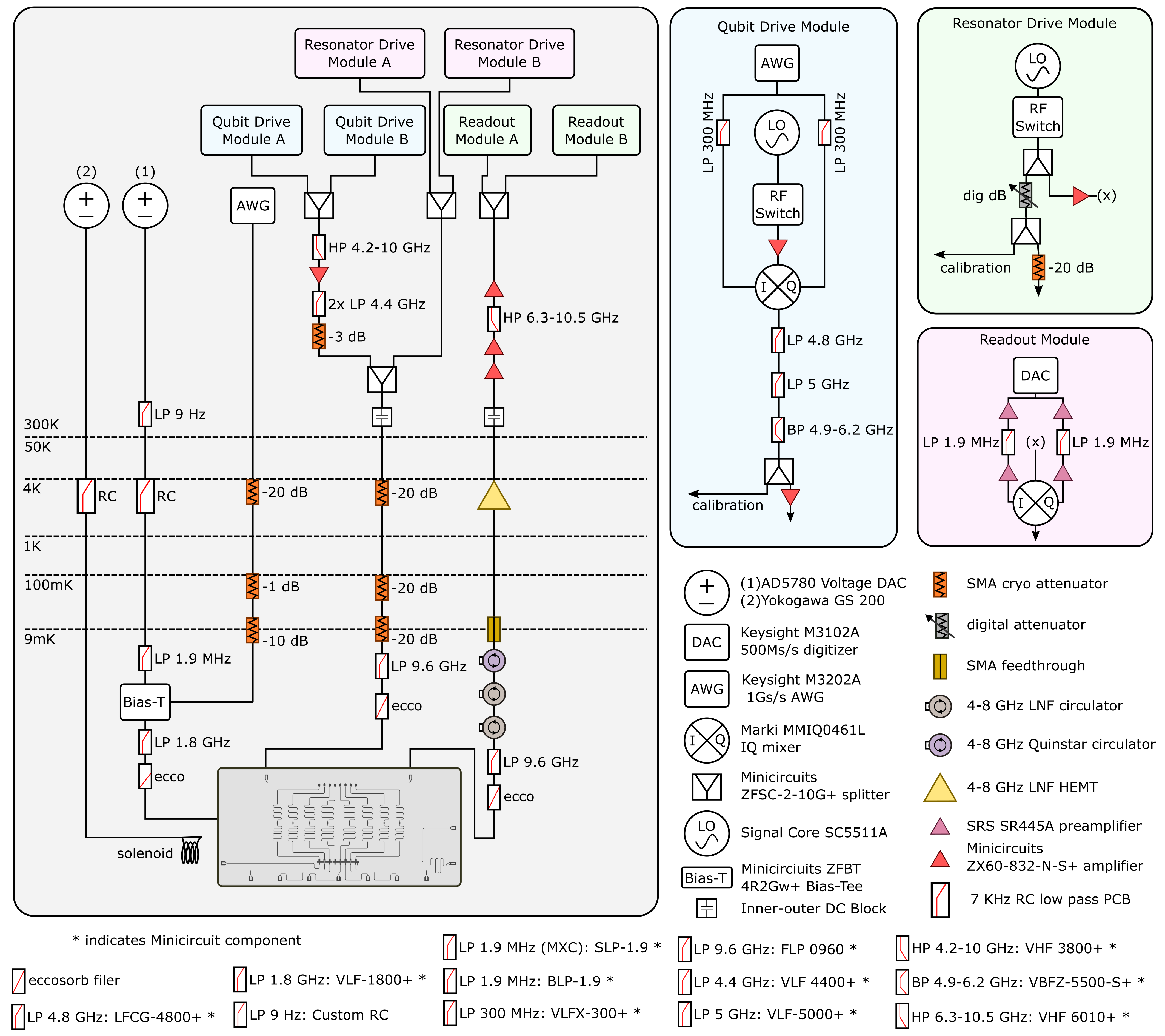}
	\caption{
		\textbf{Experiment Setup} Schematic drawing of cryogenic and instrumentation setup.
	}
	\label{fig:SI wiring}
\end{figure*}

The packaged sample is mounted to the base plate of a BlueFors dilution refrigerator at a nominal temperature of 8~mK. A solenoid of coiled niobium-titanium (NbTi) wire is fixed to the packaged sample to provide a global bias field with little heating, useful for getting close to desired lattice flux bias point without driving to much current through the DC flux lines. The sample is placed in a heat-sunk can consisting of a thin high-purity copper shim shield, followed by a high-purity superconducting lead shield, followed by two $\mu$-metal shields (innermost to outermost) to provide additional shielding from radiation and external magnetic fields.

The experimental setup is shown in~\ref{fig:SI wiring}. All input and flux lines have Eccosorb (CRS-117) filters at base to protect the sample from IR radiation. A Keysight PNA-X N5242 is used to perform spectroscopy of the readout resonators/Purcell filters and to measure the flux tuning curves of each qubit. For experiments, a SignalCore SC5511A is used to generate an LO tone near the qubit frequency. This tone is windowed using a custom-built microwave switch, then I/Q modulated by a Keysight PXIe AWG (M3202A, multichannel, 1~GS/s) to generate the individual qubit drive pulse. There are two copies of this qubit drive setup, enabling simultaneous pulses to the qubits staggered both above and below the lattice frequency, a bandwidth otherwise inaccessible with the 1~GS/s AWG. The readout tones are also generated with SC5511A SignalCores  windowed by microwave switches. There are two readout LOs, enabling simultaneous two-qubit readout. The qubit drive and readout pulses are combined outside the fridge and sent to the sample on a common input line. The output signal coming from the chip passes through three cryogenic isolators (64 dB total isolation) at base, then travels via a NbTi low-loss superconducting line to the 4K plate where it is amplified by a Low Noise Factory HEMT amplifier.The output signal coming out of the fridge passes through additional room temperature amplifiers (Miteq AFS3-00101200-22-10P-4, Minicircuits ZX60-123LN-S+). The signal is then split and demodulated against both of the readout LO tones using an I/Q mixer, and recorded using a fast digitizer (Keysight M3102A, 500 MSa/s). DC and RF pulses are combined at a MiniCircuits Bias Tee (ZFBT-4R2GW) with the capacitor removed before traveling to the chip. Each qubit has its own combined RF+DC flux line.

\section{Pulse Sequences and Operating points}
\label{SI:PulseSequencesandOperatingPoints}
This paper uses two kinds of frequency configurations for the qubits: one where the qubits are in a disordered stagger, and the other where the qubits are all degenerate. All our experiments involve tuning the qubits between these frequency configurations either diabatically or adiabatically depending on what the experiment requires. 

In the disordered stagger, three qubits are above the lattice frequency and four qubits are below. The qubits are staggered in a zig-zag, so that neighbors are detuned by much greater than $J$. This configuration is useful for both readout and state preparation. Since eigenstates consist of localized particles and are well separated in energy, eigenstate preparation simply consists of applying a sequence of calibrated microwave $\pi$-pulses. Similarly in this disordered configuration, each readout resonator is only sensitive to the population of its individual qubit, enabling site-resolved readout.

There are two versions of the disordered stagger used in this experiment. 
The first is the large disordered stagger (given in Table~\ref{SI:systemparameters}), where neighboring qubits are spaced by $>U$ apart, and next-to-nearest-neighbors are spaced by $> 2 J$. This stagger configuration is good for readout, but not for state preparation. At this lattice, we notice minimal readout crosstalk (values listed in SI Fig.~\ref{fig:SI rd xtalk}). However, since the qubits are staggered with neighbors spaced by $>U$ (i.e. the $\ket{2}$ state of the upper qubits is above the $\ket{1}$ state of the lower qubits), the highest energy eigenstate for $N>3$ particle prep involves prepping $\ket{2}$ states of the upper three qubits. This means that when ramping all the qubits to degeneracy, the ramp needs to be adiabatic both for the upper qubit $\ket{2}$ crossing the lower $\ket{1}$ as well as adiabatic when all the qubits approach degeneracy. Dealing with extra transitions is non-ideal, as it increases the ramp time needed to preserve adiabaticity. 

In the second, smaller disordered stagger configuration, neighboring qubits are spaced by $\simeq \frac{U}{2}$ apart, and next-to-nearest-neighbors are spaced by $\sim 1.5 J$. This stagger configuration is good for state preparation, but not good for readout. At this lattice, we measure more significant readout crosstalk (more than 20\% on at least four sites in any given configuration). However, eigenstates are still very localized, so preparing eigenstates still can be done with $\pi$-pulses. Further, ramping to degeneracy does not involved any $\ket{2}$ state crossings. 

For all data in this paper except for the entanglement vs disorder profiles, states were prepared in the small disordered stagger and read out in the large disordered stagger.

\begin{figure*} 
	\centering
	\includegraphics[width=0.95\textwidth]{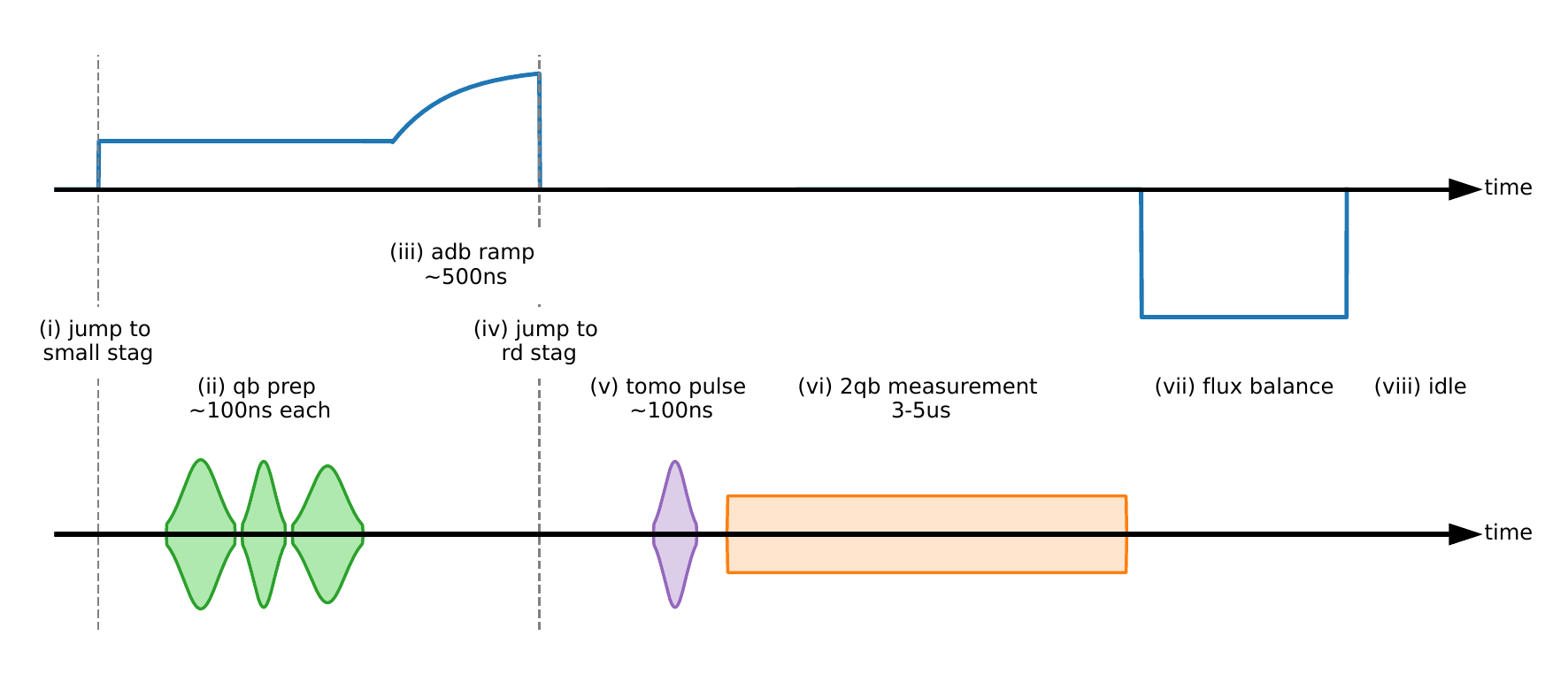}
	\caption{
		\textbf{Pulse Sequence} All experiments begin with the qubits at the large disordered stagger. First (i), qubits are jumped to the small disordered stagger using RF flux (blue). Once the qubit frequencies settle, (ii) qubit preparation pulses are applied (green). Qubits are (iii) adiabatically ramped to degeneracy, then (iv) jumped back to the large stagger for single-site-resolved readout. (v) X90 and Y90 tomography gates (purple) are applied if the experiment involves full qubit tomography. (vi) a long readout pulse (orange) is applied to the resonator/s of the qubit/s being read out. Finally, (vii) a flux balancing pulse is applied to stop long timescale inductances from building up. The experiment idles so that the total experiment time is 500 $\mu$s to ensure all the qubits decay and are reset to the $\ket{0}$ state for the start of the next experiment. 
	}
	\label{fig:SI expt seq}
\end{figure*}

A typical experiment pulse sequence is illustrated in Fig.~\ref{fig:SI expt seq}. Before the start of an experiment, qubits are tuned to the large disordered stagger using DC flux bias. All microwave signals are off and the lattice is empty. During an experiment, qubit frequencies are controlled using the RF flux lines. At the start of an experiment, qubits are (i) diabatically ``jumped'' (i.e. rapidly tuned) to the small disordered stagger configuration for state preparation. After waiting 600ns to ensure the qubit frequencies are stable, (ii) $\pi$-pulses are sequentially applied to several of the qubits. The qubits are (iii) adiabatically ramped to lattice degeneracy, adiabatically preparing the state of interest. After the qubits hit degeneracy/the state is prepared, (iv) the qubits are jumped to the large disordered stagger frequency configuration to freeze tunneling dynamics and onsite occupancy (or for reversibility experiments, ramped adiabatically back to the small stagger and then jumped to the large stagger for readout). After waiting $600$ ns for qubits to settle, (v) X90 and Y90 pulses are applied if doing tomography. Then, (vi) a readout pulse is applied (or two readout pulses are applied simultaneously if doing correlation measurements). Finally, a flux-balancing pulse is applied so that the total experiment flux is zero, to avoid charging up any stray long-term inductances~\cite{Johnson2011Flux}. The experiment is repeated every 500~$\mu$s, leaving enough idle time between pulse sequences to ensure all the qubit excitations decay and are reset to the $\ket{0}$ state for the start of the next experiment. 

We chose an exponential ramp shape for our adiabatic qubit ramp. The timing to preserve adiabaticity for different ramp shapes will obviously differ, but the exponential ramp simplifies the number of free variables to optimize over and captures the usual process of the many-body gap decreasing as more lattice sites approach resonance with one-another. For data in this paper, we found an exponential ramp of the form $e^{-t/\tau}$ gave good reversibility results (see Fig.~\ref{fig:SingleParticleKStates}) with $\tau=\frac{2}{5}\times t_{ramp}$. All qubit pulses are Gaussians truncated at $\pm 2 \sigma$.

When the system is at the small disordered stagger configuration, placing an excitation in a qubit shifts the frequency of its neighbor by $\sim 1$ MHz. This is expected from theory, and occurs because of the proximity of the upper qubits' $\omega_{12}$ transition to the lower qubits' $\omega_{01}$ transition. We take these shifts into consideration by calibrating $\pi$-pulse amplitude and frequency depending on the order of qubit preparation.

\begin{figure*} 
	\centering
	\includegraphics[width=0.95\textwidth]{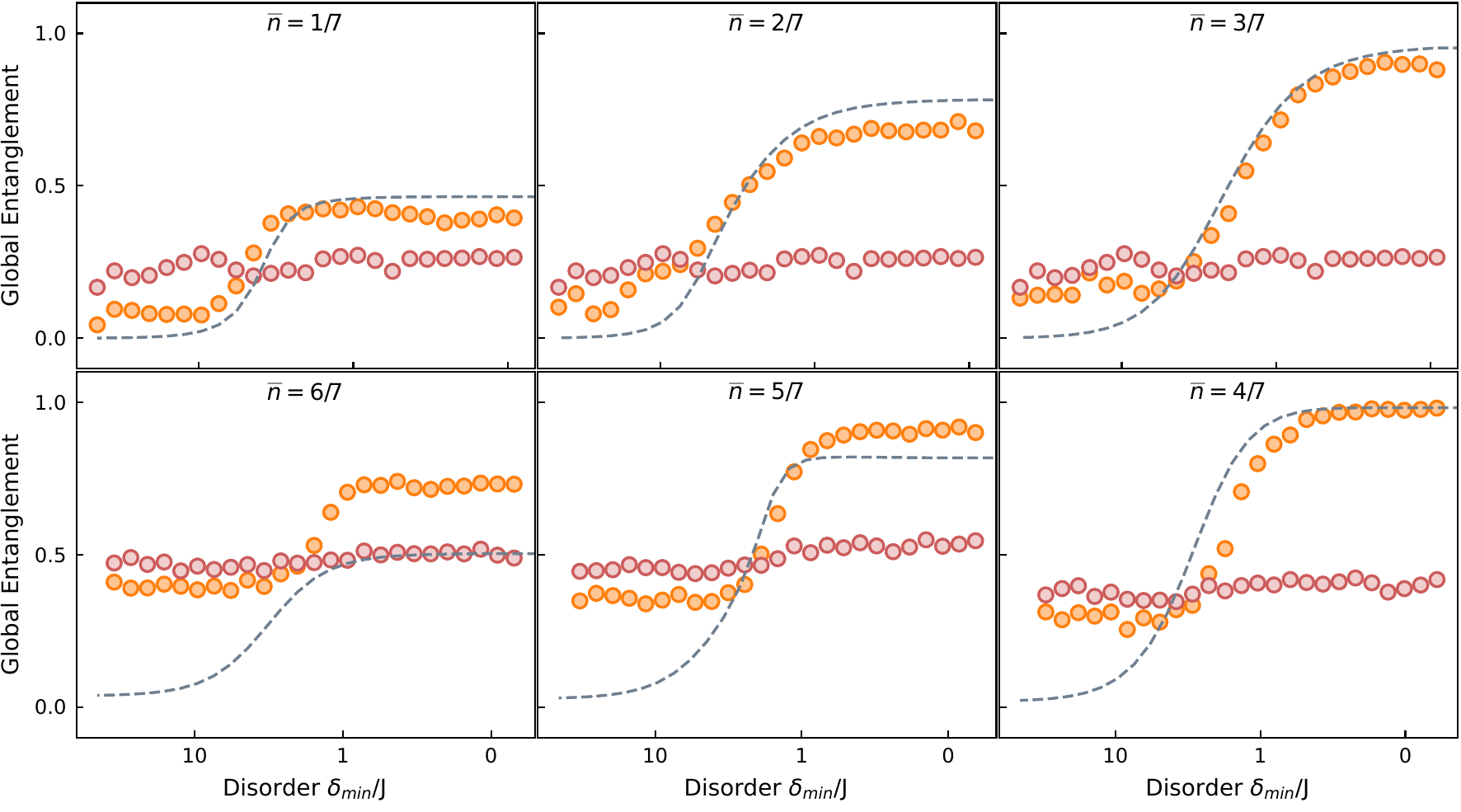}
	\caption{
		\textbf{Entanglement for $\bar{n}$ Fillings} Data: Measure of Entanglement vs Disorder, for Filling $\bar{n}$=$\frac{1}{7}$ through $\frac{6}{7}$. Error bars reflect S.E.M.; here, they are smaller than markers.
	}
	\label{fig:SI Egl all}
\end{figure*}

For entanglement vs disorder profiles, we used different preparation methods in order to measure entanglement starting at greater disorders than the small stagger. For $N<4$ entanglement vs disorder profiles (see SI~\ref{fig:SI Egl all}), the qubits are prepped at the large stagger and ramped to degeneracy directly from there. For $N\ge$4 entanglement vs disorder profiles, qubits are prepped at the large stagger, all filled qubits are then jumped to lattice degeneracy, and the holes are ramped in.  We do not use this method for $N>3$ generally because we cannot jump filled qubits directly to the lattice fast enough to be fully diabatic, so we get a small but significant number of unwanted transitions (see SI~\ref{SI:Readout}).

\section{Purcell Filter}
\label{SI:PurcellFilter}

In a system where a qubit is dispersively coupled to one readout resonator, there exists a loss channel for qubit excitations leaving the system through the measurement channel of the resonator referred to as Purcell loss. By adding another element, a Purcell filter, between the resonator and the 50-Ohm environment we can re-shape this impedance and increase the limited lifetime we would face from this loss process as we sweep our transmon qubits through frequency. To implement this simply, we add yet another linear resonator between the existing readout resonator and the environment. This new filter resonator is designed with a large linewidth centered at the readout frequency such that photons at the readout frequency pass to the measurement environment while photons at the qubit frequency are mostly reflected. In our design each qubit is capacitively coupled in series to two $\lambda /2$ CPW resonators: first to a narrow-linewidth readout resonator and then to a larger-linewidth Purcell filter resonator before capacitively connecting to the common feedline bus. We use the narrow dispersively coupled readout mode to probe the state of the qubit. To extract physical parameters (see SI~\ref{SI:systemparameters}) we perform a hanger measurement of the double-resonator system via the feedline and fit the transmission spectrum to a simplified quantum-optical model of a transmission line connected to two coupled resonators, as done in similar implementations of Purcell filters~\cite{Heinsoo2018RapidQubits}.

\section{Readout}
\label{SI:Readout}
We perform individual-site readout of our lattice using heterodyne dispersive readout (see Table~\ref{SI:systemparameters} for readout parameters, $\chi$, and fidelities). For each shot, we average the readout voltage for $5\,\mu$s, and then based on the averaged voltage value bin that shot as either the transmon $\ket{0}$, $\ket{1}$, or $\ket{2}$ state. We chose two different readout frequencies, $\omega_{\textrm{read}01}$ and $\omega_{\textrm{read}12}$. At  $\omega_{\textrm{read}01}$ the distinguishability of $\ket{0}$ from $\ket{1}$/$\ket{2}$ is maximized, while the distinguishability of $\ket{1}$/$\ket{2}$ is minimized. Similarly, at  $\omega_{\textrm{read}12}$ the distinguishability of $\ket{2}$ from $\ket{0}$/$\ket{1}$ is maximized, while the distinguishability of $\ket{0}$/$\ket{1}$ is minimized. With these optimized readout points, we measure a single shot fidelity for binning $\ket{0}$ and $\ket{1}$ at  $\omega_{\textrm{read}01}$ of $85\% - 95\%$.

With $85\% - 95\%$ fidelity, binning errors are non-negligible. To correct for this, we use a confusion matrix measured at the end of each experiment. To measure the confusion matrix, we apply a readout pulse (qubit should be in $\ket{0}$) and measure $\ket{0}$ and $\ket{1}$ counts. Then, we apply a $\pi$-pulse (qubit should be in $\ket{1}$) and measure the number of $\ket{0}$ counts and $\ket{1}$ counts. From these measurements, we can construct a $2\times 2$ confusion matrix. Inverting this matrix and applying it to our data corrects for binning errors. It also corrects for both average rates of population loss and thermal excitations during readout. For simultaneous two-qubit measurements, we measure and calibrate using a two-qubit ($4\times 4$) confusion matrix. 

There are two types of error in the experiment not corrected for by the confusion matrix: readout crosstalk and Landau-Zener transitions.

\begin{figure} 
	\centering
	\includegraphics[width=0.5\textwidth]{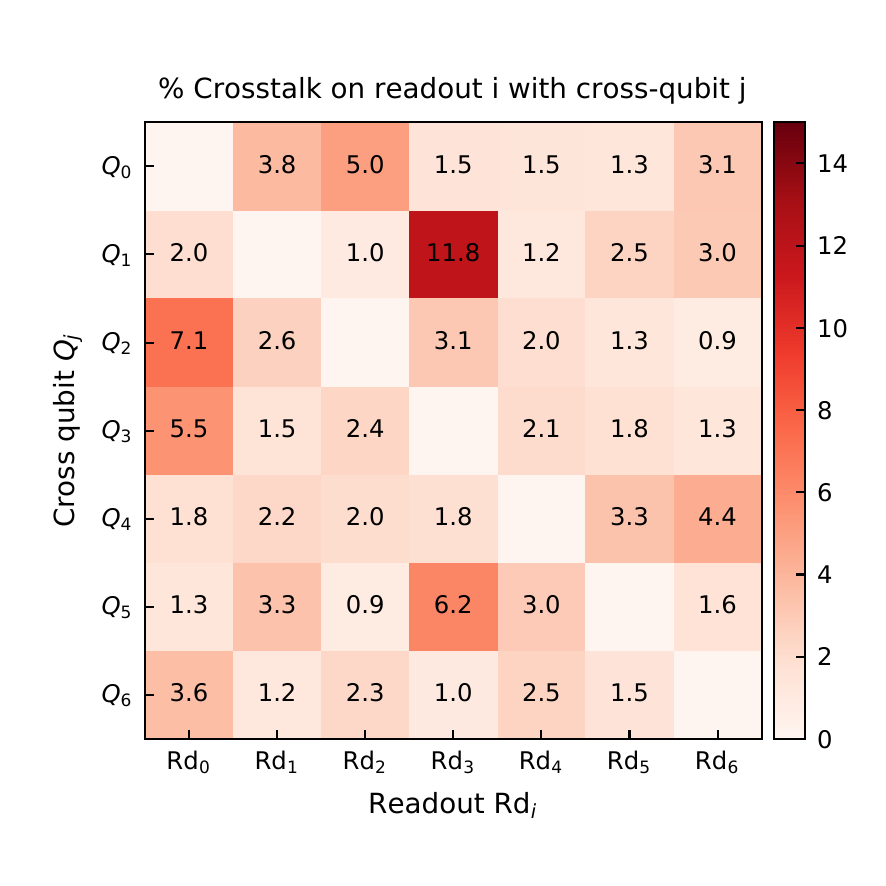}
	\caption{
		\textbf{Readout Crosstalk} Measure of how sensitive readout resonator Rd$_i$ is to the state of qubit $Q_j$, normalized to the shift of resonator Rd$_i$ in response to the state of qubit $Q_i$.
	}
	\label{fig:SI rd xtalk}
\end{figure}

Readout crosstalk occurs when a readout resonator is sensitive to a qubit other than the one it is directly coupled to, causing binning errors depending on the population of that other qubit. The source of this type of crosstalk in our system is not completely apparent. We measure that the crosstalk is worse the closer the qubits are together in frequency, but even at the large disordered stagger we were unable to find a frequency configuration without at least some readout crosstalk. At the large disordered stagger used for experiments, we measure negligible crosstalk between all elements except for between Q1 and Q3 (see SI Fig.~\ref{fig:SI rd xtalk}). We suspect this crosstalk is caused in part by a box mode coupling the resonators.

Whenever we diabatically jump qubits past each other, there can be unwanted population transfer from Landau-Zener (LZ) transitions when qubit levels cross each other. In this experiment, the level crossings of concern are $\omega_{01^Q}$ with $\omega_{01^{NN}}$, and $\omega_{01^Q}$ with $\omega_{12^{NN}}$ where NN is the nearest neighbor of qubit Q. We measure the $\omega_{01^Q}$ with $\omega_{12^{NN}}$ transition from the small disordered stagger to the large disordered stagger to be <5\% for all qubits. Since the jump from degeneracy to the large disordered stagger is bigger and thus the qubits move past each other faster, we can assume that the LZ transition for that frequency jump will also <5\% for all qubits. We calculate that the $\omega_{01^Q}$ with $\omega_{01^{NN}}$ transition probability will also be <5\% for all qubits.

\section{Effective Model as a Tonks-Girardeau Gas}
\label{SI:TGgas}
The physics of our system can be well modeled as a continuous Tonks-Girardeau (TG) gas. The TG gas describes the behavior of bosons in 1D at low temperatures, where the repulsive interactions dominate. Rather than collapsing into a simple condensate where the wavefunction is nearly a product of single-particle states, the TG boson wavefunction exhibits zeros whenever two bosons occupy the same position in space ~\cite{Paredes2004TonksGirardeauLattice, Kinoshita2004ObservationGas}. Girardeau exactly solved the TG gas problem in 1960 by mapping the 1D gas of impenetrable bosons onto a gas of noninteracting spinless fermions~\cite{Carusotto2013QuantumLight, girardeau1960relationship}.

\begin{figure*} 
	\centering
	\includegraphics[width=0.95\textwidth]{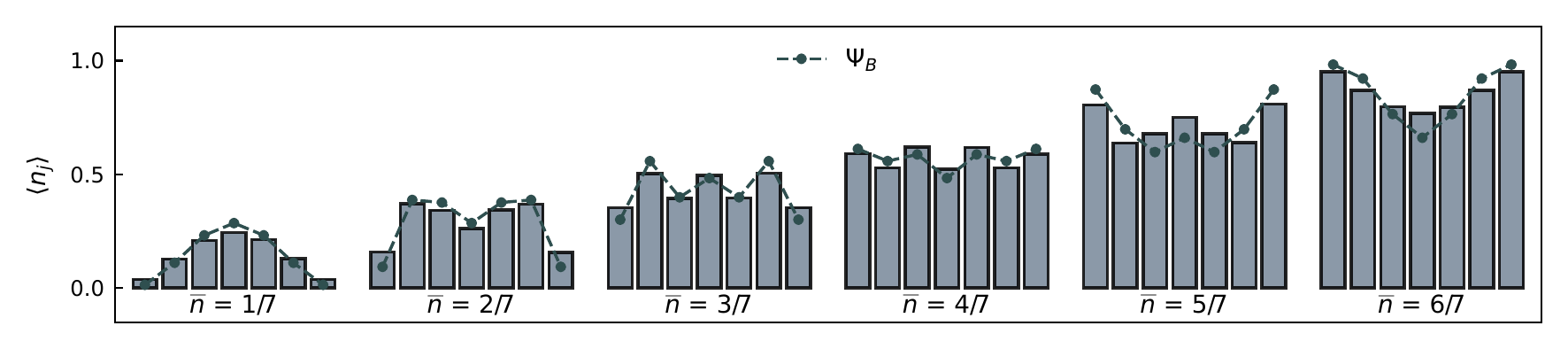}
	\caption{
		\textbf{TG Gas Wavefunction Compared to Exact Diagonalization} The bar plots correspond to the densities calculated numerically from exact diagonalization, while the dashed lines come from the analytic TG gas Bijl-Jastrow~\cite{bijl1940lowest} wavefunction. The two are in very close agreement.
	}
	\label{fig:SI TG gas wv}
\end{figure*}

In our 1D system of hardcore bosons, with U$\gg$J we are close to the TG gas regime of infinite boson repulsion. Even though our system is a discretized lattice, we find that its properties are still very well captured by a continuous TG gas model. We assume a 1D ground state many-body wave function of the Bijl-Jastrow~\cite{bijl1940lowest} form $\Psi_B(\mathbf{x}) = \phi(\mathbf{x}) \varphi(\mathbf{x})$, for $\mathbf{x} = (x_0, x_1,\ldots, x_6)$, where the open boundary condition of our lattice (i.e. the potential well) is captured by the component $\phi(\mathbf{x}) = \prod_{i=0}^{6} \cos(\pi x_i / L)$, and the two-particle component $\varphi(\mathbf{x}) = \prod_{i<j} |x_i - x_j|$ gives the TG gas impenetrable boson requirement~\cite{Rigol2011Bose1DRev}. Using this trial wavefunction, we calculate density profiles for different particle numbers in the potential, and find very close agreement between exact diagonalization of our lattice and the results from the analytic wavefunction (see SI Fig.~\ref{fig:SI TG gas wv}). 

\begin{figure} 
	\centering
	\includegraphics[width=0.5\textwidth]{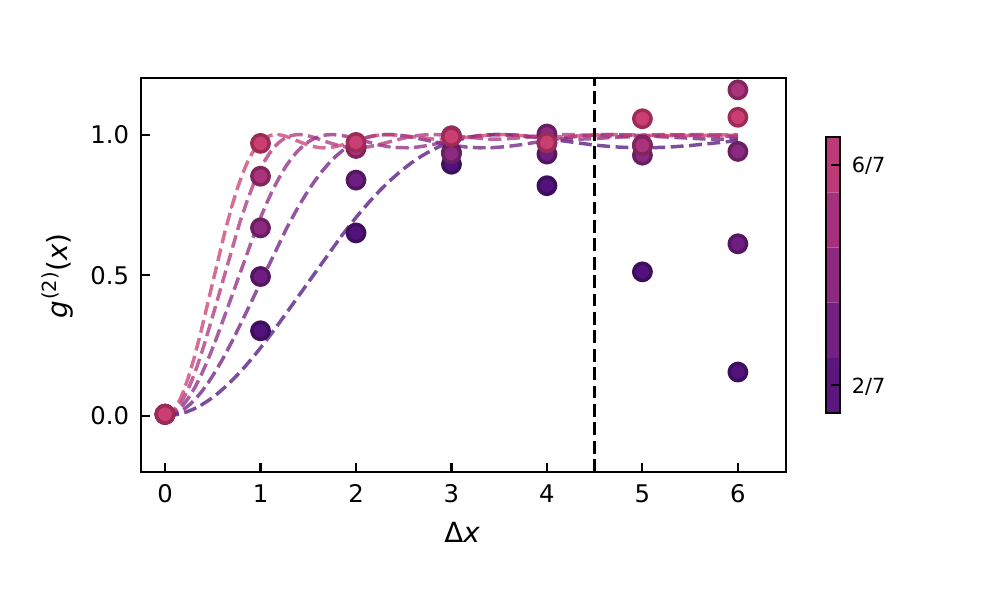}
	\caption{
		\textbf{$g^{(2)}(x)$ for $\bar{n}$ Fillings} Measure of the two-body correlator $g^{(2)}(x)$ for all lattice spacings. In the main text, we only displayed results to the left of the black dashed line. To the right of the black dashed line, our results differ from infinite 1D TG theory due to finite size effects. Error bars reflect S.E.M.; here, they are smaller than markers.
	}
	\label{fig:SI g2 all}
\end{figure}

The TG gas also very closely captures the behavior of density-density correlators in our system, as shown in SI. Fig.~\ref{fig:correlations}. Our system exhibits both particle repulsion and fermionization-induced Friedel oscillations predicted by TG gas model~\cite{Rigol2011Bose1DRev}. Note that because we have a a finite system, at larger $x$ separations our results begin to diverge from the unconstrained TG gas result; because of finite size effects and the hard open boundary of our system, correlations drop at larger separations (see SI Fig.~\ref{fig:SI g2 all}). This effect is captured in our numerics when we exactly diagonalize our system for up to 6 particles.

\section{DC and RF Flux Crosstalk Calibrations}
\label{SI:FluxCrosstalk}
The experiment requires precise control of on-site qubit frequencies $\omega_{01}$. The qubits can be tuned by applying current to their flux bias lines (and thus threading flux through their SQUID loops). We use use DC flux bias to tune to our target frequencies, and RF flux bias for ns to $\mu$s time-scale tuning during a specific experiment. The flux through a given SQUID loop for a given qubit $Q_i$ is affected by all currents $I_j$ nearby; thus, there is non-negligible flux crosstalk between the qubits. In order to be able to control each qubit's frequency individually, we need to measure and correct for both DC and RF crosstalk. We follow the same procedure as in~\cite{Ma2019AuthorPhotons}, except that we use qubit spectroscopy instead of Ramsey interferometry to measure the crosstalk matrix elements. We fit measured qubit frequency vs flux to a Jaynes-Cummings model to extract $\omega_{01}(\phi)$. DC and RF crosstalk matrices are displayed below.

The accuracy of frequency tuning is limited by the accuracy of our crosstalk measurement and the accuracy of our $\omega_{01}(\phi)$ measurement. Using the DC crosstalk matrix and the fitted $\omega_{01}(\phi)$ relation,  we are able to hit intended on-site frequencies to within $\delta_{01} \lesssim 2\pi\times 10$ MHz with no corrections, and $\delta_{01} \lesssim 2\pi\times 100$~kHz after a few rounds of local corrections to the $\omega_{01}(\phi)$ relation. Using the RF crosstalk matrix, we are able to hit intended on-site frequencies to within $\delta_{01} \lesssim 2\pi\times 2$ MHz with no corrections. To ensure we hit lattice degeneracy, we feed back on the local $\omega_{01}(\phi)$ relations by comparing our many-body profiles to expected theory. 

\begin{figure*} 
	\centering
	\includegraphics[width=0.95\textwidth]{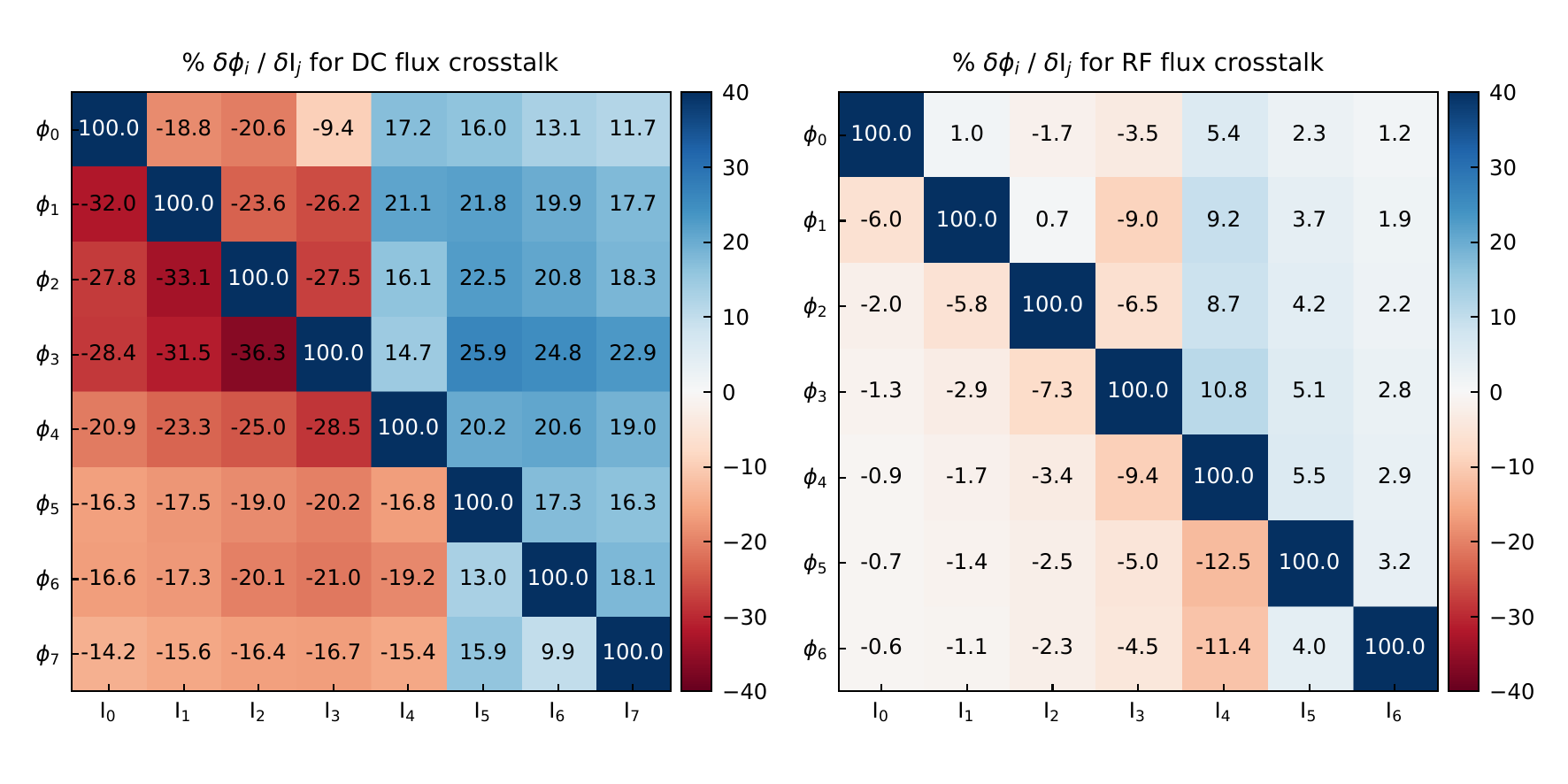}
	\caption{
		\textbf{DC and RF Flux Crosstalk Matrices}
	}
	\label{fig:SI CTMs}
\end{figure*}

\section{Flux Kernel Correction}
\label{SI:Kernel}
\begin{figure} 
	\centering
	\includegraphics[width=0.5\textwidth]{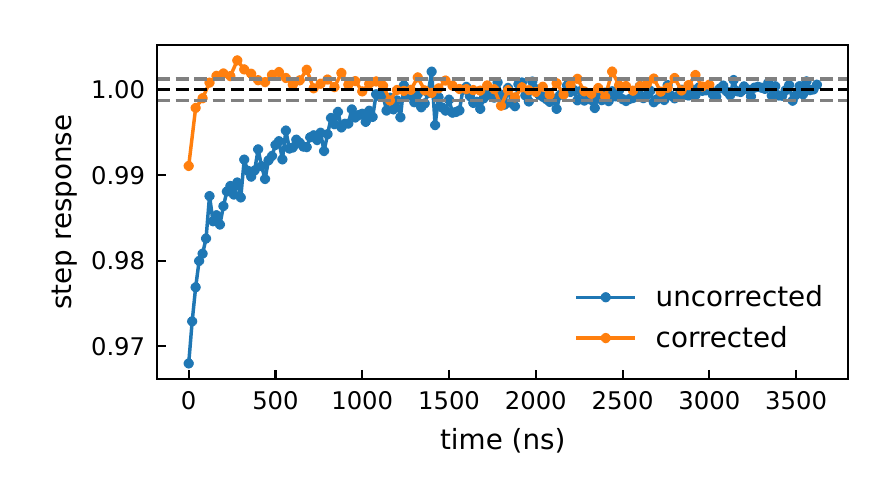}
	\caption{
		\textbf{Qubit Frequency Response to a Step Function in Flux With and Without Kernel Correction}
	}
	\label{fig:SI flux kernel}
\end{figure}

We would like to control the qubit frequencies with great precision using RF flux. However, the flux signal is distorted by low-pass filtering on the RF line (in our case, AWG distortion is negligible). We use the same procedure as in~\cite{Ma2019AuthorPhotons} to measure the distortion using qubit spectroscopy and correct for it. An example of a flux step pulse pre- and post- correction is plotted in Fig.~\ref{fig:SI flux kernel}. For the corrected signal, the qubit frequency $\omega/2\pi$ settles to within $<0.125\%$ ($0.5$~MHz for a $400$~MHz jump, which is the largest jump in our experiments) in $400$~ns.

\section{Simulations}
To verify our experimental results, we simulate properties and dynamics of this system using Qutip. In our simulations, we use a Hamiltonian of the form:
\[
\mathbf{H}_\mathrm{BH}/\hbar = -\sum_{ \langle i,j \rangle}{J_{ij}  a_i^\dagger a_j } + \frac{U}{2}\sum_i{n_i \left(n_i-1\right)} + \sum_i {\omega_i n_i} \label{eq:bosehubbardB}.
\]
plugging in experimentally measured values for $J$, $U$, and $\omega$.  We truncate our Hilbert space at the $\ket{3}$ state (3 photons) of each qubit. We operate in the energy restricted (ENR) subspace of 6 total excitations for our simulations. We numerically diagonalize the Hamiltonian in order to obtain the eigenergies and eigenstates to verify many-body state spectrum, profiles, global entanglement, and $g^{(2)}$.

In order to simulate our dynamics and obtain theory for adiabiticity times, we numerically solve the Schrodinger equation for different eigenstates of the Hamiltonian. 

\section{Uncertainty Calculations}
\label{SI:errorbars}
For each experiment, we measure 2000 shots, bin the shots, apply relevant confusion matrices, and extract the averaged quantity of interest (MB profile, entanglement, $g^{(2)}$, etc). We then repeat the experiment 10-11 times, and calculate the mean of the averages and standard deviation of the averages (i.e. calculate the standard error on the mean, or S.E.M.) for the values and error bars that we report in this paper. The experiment repetitions are performed close in time (typically within a 10-30 min span) so that our error bars are not affected by slow experimental drift over hours or days.

\section{Additional Data}
\label{SI:AdditionalData}

\begin{figure*} 
	\centering
	\includegraphics[width=0.95\textwidth]{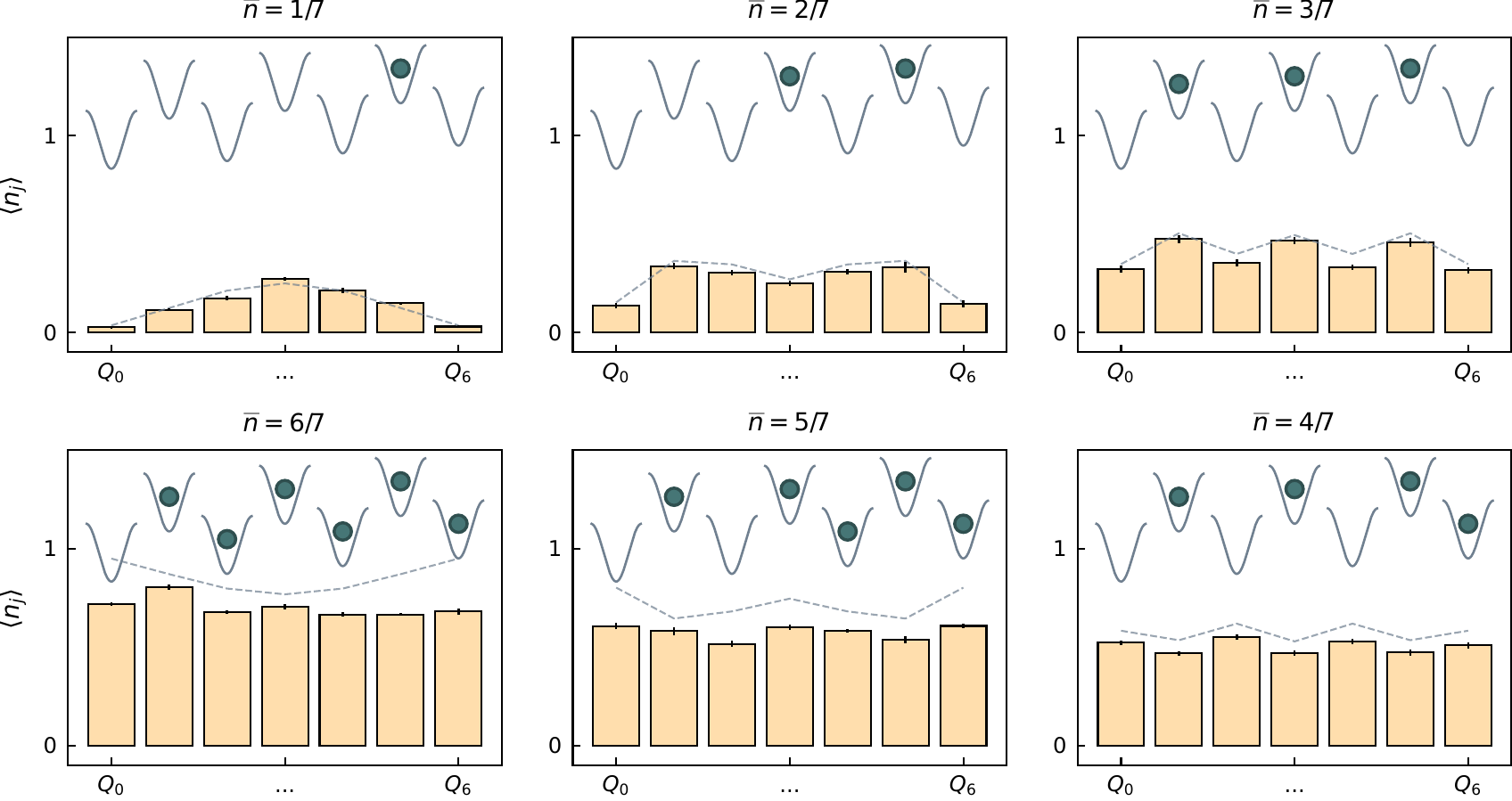}
	\caption{
		\textbf{Profiles for $\bar{n}$ Fillings} Density profiles for the highest energy eigenstates, corresponding to fluid ground states, for filling $\bar{n}$=$\frac{1}{7}$ through $\frac{6}{7}$. For 5 and 6 particles, our results suffer from particle loss.
	}
	\label{fig:SI Npart prof}
\end{figure*}

\begin{figure} 
	\centering
	\includegraphics[width=0.95\textwidth]{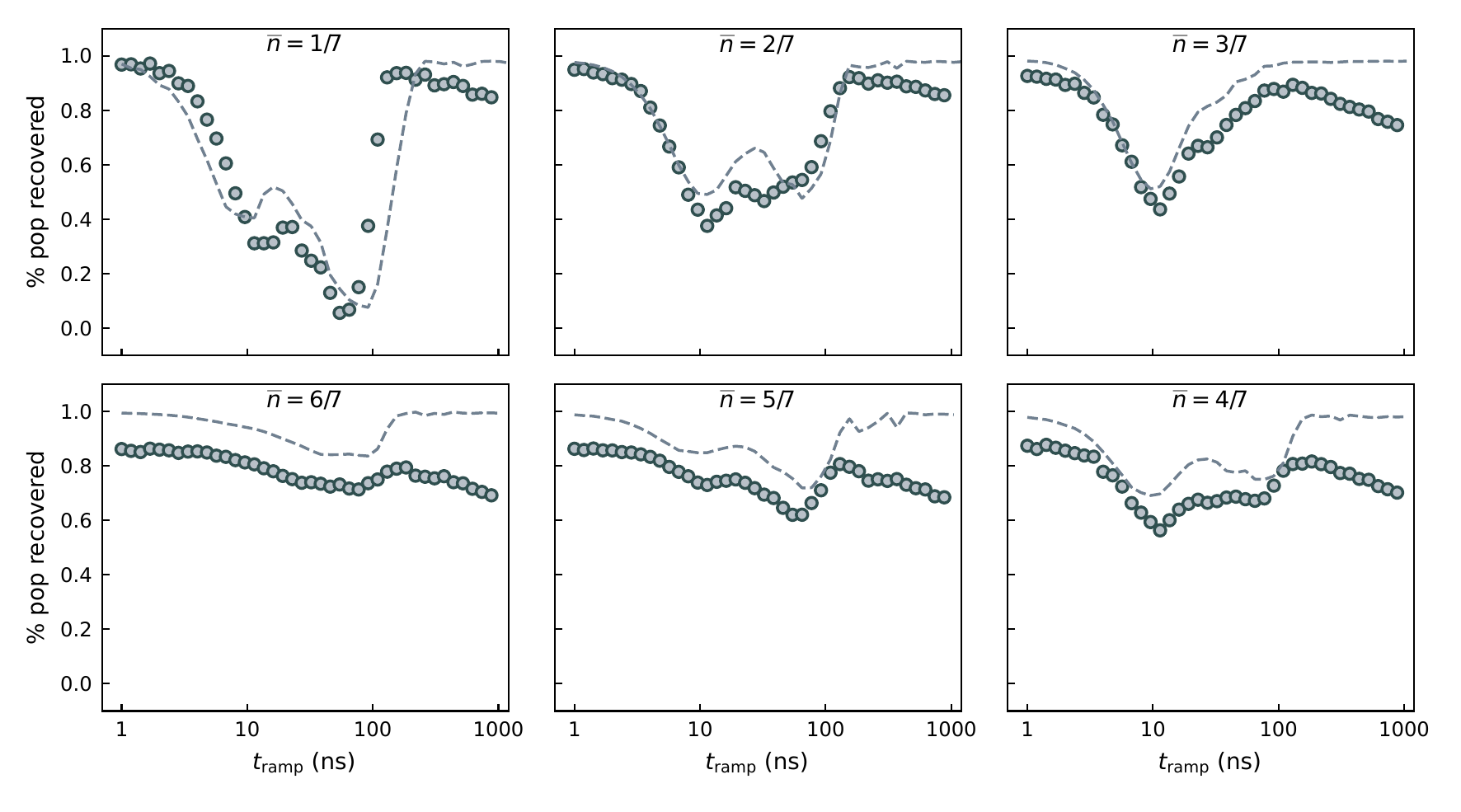}
	\caption{
		\textbf{Adiabaticity Curves for All $\bar{n}$ Fillings.} Adiabaticity is given by the average number of photons that return to the originally excited sites as a function of ramp length. Here, we measure adiabiaticty curves for the highest energy eigenstate for all $\bar{n}$ fillings, revealing the minimum ramp length needed to be adiabatic when preparing these many-body states. As particle number increases, we start to suffer more from loss and no longer fully recover the initial starting population.
	}
	\label{fig:SI adb all}
\end{figure}




\
\end{document}